\documentclass[aps,prb,twocolumn,showpacs,groupedaddress]{revtex4}
\usepackage[dvips]{graphicx}
\usepackage{amsmath}
\renewcommand{\vec}{\mathbf}
\renewcommand{\Im}{\text{ Im}}
\begin{document}
\title{Ferromagnetism in the Kondo-lattice model}
\author{C.~Santos}
\author{W.~Nolting}
\affiliation{Lehrstuhl Festk{\"o}rpertheorie, Institut f{\"u}r Physik,\\
  Humboldt-Universit{\"a}t zu Berlin, Invalidenstr.\ 110, 10115 Berlin, Germany}
\date{\today}
\begin{abstract}
        We propose a modified RKKY (Rudermann, Kittel, Kasuya, Yosida)
        technique to evaluate the magnetic properties of the
        ferromagnetic Kondo-lattice model. Together with a previously developed self-energy
        approach to the conduction-electron part of the model, we obtain a closed system of equations
        which can be solved self-consistently. The results allow us to study the conditions for 
        ferromagnetism with respect to the band occupation $n$, the interband exchange coupling $J$
        and the temperature $T$. Ferromagnetism appears for relatively low electron (hole) densities, while
        it is excluded around half-filling ($n=1$). For small $J$ the
        conventional RKKY theory ($\sim J^2$) is
        reproduced, but with strong deviations for very moderate
        exchange couplings. For not-too-small 
        $n$ a critical $J_c$ is needed to produce ferromagnetism with a finite Curie temperature
        $T_{\textrm{C}}$, which increases with $J$, then running into a kind of saturation, in order to
        fall off again and disappear above an upper critical exchange
        $J$. Spin waves show a uniform
        softening with rising temperature, and a nonuniform behaviour as functions of $n$ and $J$. The 
        disappearance of ferromagnetism when varying $T$, $J$ and $n$ is uniquely connected to the
        stiffness constant $D$ becoming negative.
\end{abstract}
\maketitle
\section{Introduction}\label{sn:sc:intro}
The Kondo-lattice model (KLM)\cite{ZEN51, AHA55, KAS56, KON64, NAG74,
NOL79, OVC91, DDN98} describes the interplay of itinerant electrons in
a partially filled conduction band with quantum-mechanical spins
(magnetic moments) localized at certain lattice sites. Typical model
properties result from an interband exchange between the two
subsystems. They are provided by a model Hamiltonian, which consists of
two parts:
\begin{equation}\label{sn:eq:Hamil}
  H=H_s+H_{sf}.
\end{equation}
$H_s$ describes uncorrelated electrons in a nondegenerate energy
band (''$s$~electrons'')
\begin{equation}\label{sn:eq:Hs}
  H_s=\sum_{ij\sigma}T_{ij}c^{\dagger}_{i\sigma}c^{}_{j\sigma},
\end{equation}
 where $c^{\dagger}_{i\sigma}(c_{i\sigma})$ creates (annihilates)
an electron specified by the lower indices. The index $i$ refers to
a lattice site $\mathbf{R}_i$, and $\sigma=\uparrow,\downarrow$ is the
spin projection. $T_{ij}$ are the hopping integrals. The second term in
Eq.~(\ref{sn:eq:Hamil}) represents an interband exchange as an intra-atomic
interaction between the itinerant electron spin
$\boldsymbol{\sigma}_i$ and the localized spin $\mathbf{S}_i$ 
(''$f$~electrons''):
\begin{equation}\label{sn:eq:Hsfa}
  H_{sf}=-J\sum_{i}\boldsymbol{\sigma}_i\cdot\mathbf{S}_i.
\end{equation} 
According to the sign of the exchange coupling $J$, a parallel ($J>0$)
or an antiparallel ($J<0$) alignment of itinerant and localized spin is
favored with remarkable differences in the physical properties. We
restrict our considerations in the following to the $J>0$ case, sometimes
referred to as ''ferromagnetic Kondo-lattice model'', and also known as 
''$s-f$ ($s-d$) model''.
The applications of the KLM are manifold. In the 1970s it was
extensively used to describe the electronic and magneto-optic
properties of magnetic semiconductors such as the ferromagnetic EuO and
EuS \cite{NOL79, OVC91}.
Most studies focussed on the spectacular
temperature dependence of the unoccupied band states.
The well-known ''red shift'' of the optical absorption edge upon cooling
from $T=T_{\textrm{C}}$ to $T=0$ K\cite{BJW64} can be understood as a
corresponding temperature shift of the lower $5d$ conduction-band
edge. Recent quasiparticle band-structure calculation for bulk
EuO\cite{SNO01a} confirm the observed striking temperature dependence as
a consequence of a ferromagnetic exchange $J$ of the order of some tenth
an eV. The interest in EuO, in particular in thin films, was
strongly revived by the observation of a temperature driven metal-insulator
transition below $T_{\textrm{C}}$ with a large drop of the
resistivity by as much as eight orders of magnitude\cite{STE01}.
This gives rise to a colossal magnetoresistance (CMR) much stronger than
that of the manganites as La$_{1-x}$Sr$_x$MnO$_3$. Recently, 
a surface half-metal-insulator transition was
predicted\cite{SNO01b} for a Eu(100) film. The reason for this is an
exchange-split surface state. 

In ferromagnetic local moment metals like Gd a RKKY-type interaction  
is believed to cause the ferromagnetic
order, so that in principle a self-consistent description of magnetic
and electronic properties by the KLM is possible. The $T=0$-moment of Gd
is found to be $7.63\:\mu_{\textrm{B}}$\cite{RCM75}. $7\:\mu_{\textrm{B}}$
stems from the exactly half-filled $4f$ shell. The excess moment of
$0.63\:\mu_{\textrm{B}}$ is due to an induced spin polarization of the 
''a priori'' nonmagnetic ($5d,6s$) conduction-bands, indicating a
weak or intermediate $J>0$\cite{REN99}. 

An interesting application of the KLM refers to semimagnetic
semiconductors, where randomly distributed Mn$^{2+}$ or Fe$^{2+}$ ions
provide localized magnetic moments which influence, via the exchange
coupling $J$, the band states of systems like 
Ga$_{1-x}$Mn$_x$As\cite{OHN98}. The Mn$^{2+}$ ions provide both
localized spins ($S=\frac{5}{2}$, concentration $x$) and free carriers
(holes, concentration $x^{\star}<x$). The latter take care for an
indirect coupling between the moments leading for small $x$ even to a
ferromagnetic order. The highest $T_{\textrm{C}}$ so far reached is
$110$ K for $x=0.053$\cite{MOS98} where $x^{\star}$ is found to be about
$15\%$ of the Mn concentration $x$. The origin of ferromagnetism in
Ga$_{1-x}$Mn$_x$As is controversial. The authors of Ref.~\onlinecite{MOS98}
claimed that the measured $T_{\textrm{C}}(x)$ is consistent with a
conventional RKKY interaction between the Mn ions. However, they estimated
the exchange coupling $J$ to be of several eV, therewith comparable to
the Fermi energy. This makes second order perturbation theory with respect
to $J$ rather questionable. RKKY should be modified to account for 
higher-order spin polarization terms. In this papar we are going to
present a respective proposal.

Another modern application of the $J>0$ KLM aims at the manganese oxides
with perovskite structures T$_{1-x}$D$_x$MnO$_3$ (T=La, Pr, Nd;
D=Sr,Ca, Ba, Pb) which have attracted great scientific interest because
of their CMR\cite{JTM94, RAM97}.
Parent compounds like the protagonist La$^{3+}$Mn$^{3+}$O$_{3}$ are
antiferromagnetic insulators with a localized Mn spin of
$S=2$. Replacing a trivalent La$^{3+}$ ion by a divalent earth-alkali ion
(Ca$^{2+}$) leads to a homogenuous valence mixture of the manganese ion
(Mn$^{3+}_{1-x}$, Mn$^{4+}_{x}$). The three $3d-t_{2g}$ electrons of
Mn$^{4+}$ can be considered as almost localized, forming an
$S=\frac{3}{2}$ spin. Furthermore, ($1-x$) electron per Mn site are of
$3d-e_{g}$ type, and itinerant. They are exchange coupled to the
$S=\frac{3}{2}$ spins, realizing just the situation which is modelled by
the KLM. The exchange coupling is ferromagnetic ($J>0$). Since the 
manganites are bad electrical
conductors, it can be assumed that the intraatomic exchange $J>0$ is
much larger than the hopping-matrix element $|t|$. With estimated
$e_g$ bandwidths of $1-2$ eV\cite{SPV96} and lower limits for $J$ of
$\begin{array}{c}>\\[-1.5ex]\sim\\[-0.5ex]\end{array}1$ eV\cite{SPV96}, 
the manganites surely belong to the group of 
strongly coupled materials. Many fascinating features of the CMR materials
can be traced back to an intimate correlation between magnetic and
electronic components. One of the not yet fully understood features concerns
the spin dynamics in the ferromagnetic phase ($x\sim 0.3$ in
La$_{1-x}$(Ca,Pb)$_x$MnO$_3$). It was reported\cite{PAH96} that the spin
wave dispersion (SWD) of materials with high $T_{\textrm{C}}$ (e.g.:
La$_{0.7}$Pb$_{0.3}$MnO$_{3}$: $T_{\textrm{C}}=355 K$) can be
reasonably reproduced by a simple Heisenberg model with
nearest-neighbour
exchange, only. Furukawa\cite{FUR96} gave qualitative
arguments that this can be understood for strongly exchange coupled
materials. However, strong deviations from the typical Heisenberg
behavior of the SWD have been found for manganites with lower
$T_{\textrm{C}}$'s\cite{HDC98,FDH98,DHF00}. It is a challenging question
whether or not the softening of the SWD at the zone boundary can be
explained by a better treatment of the KLM, i.e.~by an appropriate
consideration of the influence of the conduction-electron self-energy on
the effective coupling between the localized magnetic
moments\cite{WAN98,VSN01}. Solovyev and Terakura\cite{STE99} argue in
such a way claiming that the SWD-softening has a purely magnetic
background. By contrast, Mancini et~al.\cite{MPP00} believe that the
KLM [Eq.~(\ref{sn:eq:Hamil})] has to be extended by a direct (superexchange)
interaction between the Mn moments to account for antiferromagnetic
tendencies, which dominate the magnetic behaviour of the parent material
LaMnO$_3$.

The above-presented, of course incomplete, list documents the rich
variety of, at least, qualitative applications for the KLM.
However, the model Hamiltonian [Eq.~(\ref{sn:eq:Hamil})] provokes such a
complicated many-body problem that approximations have to be
tolerated. Most of the recent theoretical work on the KLM, aiming at the
CMR-materials , assumed classical spins\cite{FUR99,MLS95,HVO00}, mainly in
order to apply ''dynamical mean field theory'' (DMFT). It is surely not
unfair to consider this assumption as rather problematic. 

It is the intention of the present paper to use a special RKKY technique
for a detailed investigation of the magnetic properties of the KLM with
ferromagnetic interband exchange. We exploit the method of
Ref.~\onlinecite{NRM97} to map the interaction operator
$H_{sf}$ [Eq.~(\ref{sn:eq:Hsfa})] to the Heisenberg model Hamiltonian. This is
done by averaging out the conduction electron degrees of freedom by
properly defined ''restricted'' Green functions. The resulting effective
Heisenberg exchange integrals are functionals of the itinerant-electron
self-energy, and therewith strongly temperature and band occupation
dependent. These dependencies manifest themselves in fundamental magnetic
properties such as phase diagrams, Curie temperature, spin-wave
dispersions and so on.

The electronic self-energy is taken from a ''moment conserving decoupling
approach'' (MCDA) introduced in Ref.~\onlinecite{NRM97} and
successfully applied to several topics in previous papers
\cite{SNO01a, SNO01b, REN99, SMN01}. The method carefully interpolates
between nontrivial limiting cases.

The final goal of our effort is to obtain quantitative information about
magnetism and the temperature-dependent electronic structure of real
local-moment systems. For this purpose it is intended in a future work
to combine our model study with a ''first principles'' band-structure
calculation. First attempts in this direction were done for
EuO\cite{SNO01a} and Gd\cite{REN99}; however, the magnetic ordering could
not yet be derived self-consistently.
The paper is organized as follow. In Sec.~\ref{sec:KLM} we briefly formulate the
many-body problem of the KLM. Section~\ref{sec:EffectiveE}
reproduces the most important aspects of the 
''modified RKKY-interaction'' (M-RKKY), the understanding
of which is vital for a correct interpretation of the results, which are
discussed in Sec.~IV.
\section{Kondo-lattice model}\label{sec:KLM}
The model Hamiltonian [Eq.~(\ref{sn:eq:Hamil})] provokes a nontrivial
many-body problem, which can rigorously be solved only for a small
number of limiting cases. For practical reasons, it sometimes appears
convenient to use the second quantized form of the exchange interaction
[Eq.~(\ref{sn:eq:Hsfa})],
\begin{equation}\label{sn:eq:Hsf2quant}
  H_{sf}=-\frac{1}{2}J\sum_{i\sigma}(z_{\sigma}S^{z}_{i}n^{}_{i\sigma}
         +S^{\sigma}_{i}c^{\dagger}_{i-\sigma}c^{}_{i\sigma}),
\end{equation}
with the abbreviations:
\begin{equation}\label{sn:eq:abbrevs}
  n^{}_{i\sigma}=c^{\dagger}_{i\sigma}c^{}_{i\sigma},\:
  z_{\sigma}=\delta_{\sigma\uparrow}-\delta_{\sigma\downarrow},\:
  S^{\sigma}_{i}=S^{x}_{i}+iz_{\sigma}S^{y}_{i}.
\end{equation}
The first term in Eq.~(\ref{sn:eq:Hsf2quant}) describes an Ising-like
interaction between the $z$ components of the localized and the
itinerant spin. The second term comprises spin-exchange processes between the
two subsystems.

To study the conduction-electron properties we use the single-electron
Green function,
\begin{equation}\label{sn:eq:SingleGreen}
  G_{ij\sigma}(E)=\bigl\langle\!\bigl\langle
  c^{}_{i\sigma};c^{\dagger}_{j\sigma}\bigr\rangle\!\bigr\rangle_{E}.
\end{equation}
Its equation of motion reads
\begin{eqnarray}\label{sn:eq:EQM_SG}
  \lefteqn{\sum_{m}\left(E\delta_{im}-T_{im}\right)G_{mj\sigma}(E)=\hbar\delta_{ij}-}\\\nonumber
  &&\hspace{5em}-\frac{1}{2}J\{z_{\sigma}I_{ii,j\sigma}(E)+F_{ii,j\sigma}(E)\},
\end{eqnarray}
where the two types of interaction terms in Eq.~(\ref{sn:eq:Hsf2quant}) are
causing the ''spinflip function''
\begin{equation}\label{sn:eq:spinflipfunc}
  F_{im,j\sigma}(E)=\bigl\langle\!\bigl\langle
  S^{-\sigma}_ic^{}_{m-\sigma};c^{\dagger}_{j\sigma} 
  \bigr\rangle\!\bigr\rangle_{E}
\end{equation}
and the ''Ising function''
\begin{equation}\label{sn:eq:isingfunc}
  I_{im,j\sigma}(E)=\bigl\langle\!\bigl\langle
  S^{z}_ic^{}_{m\sigma};c^{\dagger}_{j\sigma} 
  \bigr\rangle\!\bigr\rangle_{E}.
\end{equation}
These two ''higher'' Green functions prevent a direct solution of the
equation of motion (\ref{sn:eq:EQM_SG}). A formal solution of the
Fourier-transformed single-electron Green function, 
\begin{equation}\label{sn:eq:FourierSG}
  G_{\vec{k}\sigma}(E)=\frac{1}{N}\sum_{ij}G_{ij\sigma}(E)
  \:e^{i\vec{k}\cdot(\vec{R}_i-\vec{R}_j)},
\end{equation}
defines the complex electron self-energy $\Sigma_{\vec{k}\sigma}(E)$:
\begin{equation}\label{sn:eq:defselfenergy}
  G_{\vec{k}\sigma}(E)=\frac{\hbar}{E+\mu+i0^+
  -\varepsilon(\vec{k})-\Sigma_{\vec{k}\sigma}(E)},
\end{equation}
\begin{equation}\label{sn:eq:defselfcom}
\bigl\langle\!\bigl\langle[c^{}_{\vec{k}\sigma},H_{sf}]_-;c^{}_{\vec{k}\sigma}
\bigr\rangle\!\bigr\rangle_{E}=\Sigma_{\vec{k}\sigma}(E)\:G_{\vec{k}\sigma}(E).
\end{equation}
$\varepsilon(\vec{k})$ are the Bloch energies:
\begin{equation}\label{sn:eq:blochenergy}
  \varepsilon(\vec{k})=\frac{1}{N}\sum_{ij}T_{ij}
  \:e^{i\vec{k}\cdot(\vec{R}_i-\vec{R}_j)}.
\end{equation}
By use of the quasiparticle density of states 
\begin{equation}\label{sn:eq:QDOS}
  \rho_{\sigma}(E)=-\frac{1}{\hbar\pi N}\sum_{\vec{k}}\Im G_{\vec{k}\sigma}(E-\mu),
\end{equation}
we can calculate the spin-dependent occupation numbers
\begin{equation}\label{sn:eq:occnumber}
\langle n_{\sigma}\rangle=\int\limits^{+\infty}_{-\infty}dE\:f_-(E)\:\rho_{\sigma}(E),
\end{equation}
important to fix the band polarization 
$m=\langle n_{\uparrow}\rangle-\langle n_{\downarrow}\rangle$. 
$f_-(E)=\left(\exp(\beta(E-\mu)+1\right)^{-1}$ is the Fermi function.

The central quantity is the self-energy, the determination of which
solves the problem. However,  for finite temperatures and arbitrary band
occupations an exact expression is not avaliable. In
Ref.~\onlinecite{NRM97} a ''moment conserving decoupling approach''
(MCDA) predicts the following structure of the self-energy.
\begin{equation}\label{sn:eq:MCDA}
  \Sigma_{\vec{k}\sigma}(E)=-\frac{1}{2}Jz_{\sigma}\langle S^z\rangle
  +\frac{1}{4}J^2\:D_{\vec{k}\sigma}(E).
\end{equation}
The first term is linear in the coupling $J$, and proportional to the the
local-moment magnetization $\langle S^z\rangle$. It represents the
result of a mean-field approach which is correct in the weak coupling limit.
The second term is predominantly determined by spin-exchange processes
between band electrons and localized moments. It is a complicated
functional of the self-energy itself. Thus Eq.~(\ref{sn:eq:MCDA}) is not
an analytical solution at all, but an implicit equation for
$\Sigma_{\vec{k}\sigma}(E)$. The MCDA, the detail of which are presented
in ref.~\onlinecite{NRM97}, is a nonperturbational decoupling approach
to a set of properly defined Green functions that correctly reproduces
the non-trivial exact limiting cases of the KLM.  A weighty example is the
special case of a ''ferromagnetically saturated semiconductor''
\cite{SMA81,CMI01,BJW64}. The various ''higher'' Green functions have been
approximated by linear combination of ''lower'' functions, where the
rigorous spectral representations of these ''higher'' Green functions
and their exact special cases ($S=1/2$, ferromagnetic saturation, \dots)
justify the respective ansatz. Free
parameters are finally fitted to exactly known spectral moments. In a
certain sense, the method can be considered an interpolation scheme
between important limiting cases. It was successfully used in the
past for several
applications\cite{SNO01a,SNO01b,REN99,NRM97,SMN01,NMR96}. We refer the
reader for technical details to one of these papers.

The term $D_{\vec{k}\sigma}(E)$ in Eq.~(\ref{sn:eq:MCDA}) contains certain
expectations values such as 
\begin{itemize}
\item[(a)]$ \langle n_{i\sigma}\rangle,\ldots$,
\item[(b)]$ \langle S^z_i\rangle,\:\langle
  S^{\pm}_{i}S^{\mp}_{i}\rangle,\: \langle (S^z_i)^2\rangle$,\ldots,
\item[(c)]$\Delta_{\sigma}=\langle S^z_{i}n^{}_{i\sigma}\rangle,\:
  \gamma_{\sigma}=\langle S^{-\sigma}_{i}c^{\dagger}_{i\sigma}c^{}_{i-\sigma}\rangle$,\ldots.
\end{itemize}
By use of the spectral theorem, terms (a) and (c) can exactly be
expressed by the Green functions (\ref{sn:eq:SingleGreen}),
(\ref{sn:eq:spinflipfunc}) and (\ref{sn:eq:isingfunc}). Besides
Eq.~(\ref{sn:eq:occnumber}) it holds
\begin{eqnarray}
\label{sn:eq:gamma}
\gamma_{\sigma}&=&-\frac{1}{\hbar\pi}\int\limits^{+\infty}_{-\infty}dE\:f_-(E)
\Im F_{ii,i\sigma}(E-\mu),\\
\label{sn:eq:delta}
\Delta_{\sigma}&=&-\frac{1}{\hbar\pi}\int\limits^{+\infty}_{-\infty}dE\:f_-(E)
\Im \Gamma_{ii,i\sigma}(E-\mu).
\end{eqnarray} 
It remains to express the pure ''local-moment'' correlations (b) in
terms of the electronic self-energy $\Sigma_{\vec{k}\sigma}(E)$ to get a
closed system of equations that can be solved self-consistently. A finite
coupling between the spin operators must be of indirect nature since
the model Hamiltonian [Eq.~(\ref{sn:eq:Hamil})] does not contain any direct
exchange. To obtain the pure spin correlations (b), we map the exchange
interaction (\ref{sn:eq:Hsf2quant}) on an effective Heisenberg model
with effective exchange integrals $\hat J_{ij}$ being functionals of the
electronic self-energy. The result is a ''modified'' RKKY interaction
(M-RKKY) that takes into account the exchange-induced spin
polarization of the band electrons. In lowest order it reproduces the
conventional RKKY method.
\section{Effective Exchange Operator}\label{sec:EffectiveE}
The starting idea is to map the interband ($s-f$) exchange on an effective
spin Hamiltonian of the Heisenberg type.
\begin{equation}\label{sn:eq:Hf}
  H_{f}=-\sum_{ij}\hat J_{ij}\:\vec{S}_i\cdot\vec{S}_j.
\end{equation}
For this purpose we use the $s-f$ interaction in the folowing form:
\begin{equation}\label{sn:eq:Hsfk}
  H_{sf}=-J\frac{\hbar}{N}\sum_{i\sigma\sigma^{\prime}}\sum_{\vec{k}\vec{q}}
  e^{-i\vec{q}\cdot\vec{R}_i}\left(\vec{S}_i\cdot
  \boldsymbol{\hat\sigma}\right)_{\sigma\sigma^{\prime}}
  c^{\dagger}_{\vec{k}+\vec{q}\sigma}c^{}_{\vec{k}\sigma^{\prime}}.
\end{equation}
$\boldsymbol{\hat\sigma}$ is the band electron spin operator, the components
of which are Pauli spin matrices. The above-mentioned mapping occurs by
averaging out the band electron degrees of freedom:
\begin{equation}\label{sn:eq:map}
  H_{sf}\longrightarrow\langle H_{sf}\rangle^{(c)}\equiv H_f.
\end{equation}
Averaging only in the band electron subspace means that $\langle
H_{sf}\rangle^{(c)}$ retains operator character in the $f$-spin
subspace. According to Eq.~(\ref{sn:eq:Hsfk}) we have to calculate
\begin{equation}\label{sn:eq:averc+cs}
  \bigl\langle
    c^{\dagger}_{\vec{k}+\vec{q}\sigma}c^{}_{\vec{k}\sigma^{\prime}}\bigr\rangle^{(c)}  
  =\frac{1}{\Xi^{\prime}}\textrm{Tr}\left(e^{-\beta H^{\prime}}
  c^{\dagger}_{\vec{k}+\vec{q}\sigma}c^{}_{\vec{k}\sigma^{\prime}}\right).
\end{equation}
$H^{\prime}$ has exactly the same structure as the KLM Hamiltonian $H$
[Eq.~(\ref{sn:eq:Hamil})],
except for the fact that for the averaging procedure the $f$-spin
operators are to be considered as c-numbers, therefore not affecting the
trace. $\Xi^{\prime}$ is the corresponding grand partition function. The
expectation value [Eq.~(\ref{sn:eq:averc+cs})] does not necessarily vanish for
$\vec{q}\ne 0$ and for $\sigma\ne\sigma^{\prime}$, as it would do when
averaging in the full Hilbert space of the KLM. To calculate
Eq.~(\ref{sn:eq:averc+cs}) we introduce a proper ''restricted'' Green
function.
\begin{equation}\label{sn:eq:rectGreen}
\hat G^{\sigma^{\prime}\sigma}_{\vec{k},\vec{k}+\vec{q}}(E)
=\bigl\langle\!\bigl\langle c^{}_{\vec{k}\sigma^{\prime}};
 c^{\dagger}_{\vec{k}+\vec{q}\sigma}\bigr\rangle\!\bigr\rangle^{(c)}_{E},
\end{equation}
which has the ''normal'' definition of a retarted Green function, only
the averages have to be done in the Hilbert space of $H^{\prime}$. The
equation of motion is readily derived 
for both cases, namely, in the case that $c_{\vec{k}\sigma^{\prime}}$ is the
active operator or $c^{\dagger}_{\vec{k}+\vec{q}\sigma}$.
By use of the ''free'' Green function,
\begin{equation}\label{sn:eq:freeGreen}
  G^{(0)}_{\vec{k}}(E)=\frac{\hbar}{E+\mu-\varepsilon(\vec{k})},
\end{equation}
we can combine both equations:
\begin{eqnarray}\label{sn:eq:combined}
\lefteqn{\hat
  G^{\sigma^{\prime}\sigma}_{\vec{k},\vec{k}+\vec{q}}(E)=\delta_{\sigma\sigma^{\prime}}
\delta_{\vec{q},\vec{0}}\:G^{(0)}_{\vec{k}}(E)-}\\\nonumber
&&-\frac{J}{2N}\!\!\sum_{i\vec{k}^{\prime}\sigma^{\prime\prime}}
\left\{e^{-i(\vec{k}-\vec{k}^{\prime})\cdot\vec{R}_i}G^{(0)}_{\vec{k}}(E)
\left(\vec{S}_i\cdot\boldsymbol{\hat\sigma}\right)_{\sigma^{\prime}\sigma^{\prime\prime}}
\hat
G^{\sigma^{\prime\prime}\sigma}_{\vec{k}^{\prime},\vec{k}+\vec{q}}(E)\right.\\\nonumber
&&\hspace{4em}\left.+e^{-i(\vec{k}^{\prime}-(\vec{k}+\vec{q}))\cdot\vec{R}_i}G^{(0)}_{\vec{k}+\vec{q}}
\left(\vec{S}_i\cdot\boldsymbol{\hat\sigma}\right)_{\sigma^{\prime\prime}\sigma}\hat
G^{\sigma^{\prime}\sigma^{\prime\prime}}_{\vec{k},\vec{k}^{\prime}}(E)\right\}.
\end{eqnarray}
This equation is still exact and can be iterated up to any desired
accuracy.

In first order, where the restricted Green function on the right-hand
side of [Eq.~(\ref{sn:eq:combined})] is replaced by the ''free'' Green function
Eq.~(\ref{sn:eq:freeGreen}), one obtains
\begin{eqnarray}\label{sn:eq:1stOrder}
  \lefteqn{\left( \hat
      G^{\sigma^{\prime}\sigma}_{\vec{k},\vec{k}+\vec{q}}(E)\right)^{(1)}
     =\delta_{\sigma\sigma^{\prime}}\delta_{\vec{q},\vec{0}}-}\\\nonumber
&&\hspace{4em}-\frac{J}{N}\sum_{i}e^{i\vec{q}\cdot\vec{R}_i}G^{(0)}_{\vec{k}}(E)
\left(\vec{S}_i\cdot\boldsymbol{\hat\sigma}\right)_{\sigma^{\prime}\sigma}
G^{(0)}_{\vec{k}+\vec{q}}.
\end{eqnarray}
Exploiting the spectral theorem in the restricted Hilbert space, for the
expectation value [Eq.~(\ref{sn:eq:averc+cs})] one finds
\begin{multline}\label{sn:eq:averc+c1stOrder}
  \left(\frac{1}{N}\sum_{\vec{k}}\langle
    c^{\dagger}_{\vec{k}+\vec{q}\sigma}
   c^{}_{\vec{k}\sigma^{\prime}}\rangle^{(c)}\right)^{(1)}
 =\delta_{\sigma\sigma^{\prime}}\delta_{\vec{q},\vec{0}}
\frac{1}{N}\sum_{\vec{k}}f_-(\varepsilon(\vec{k}))\\
-\frac{J}{\hbar}\frac{1}{N}\sum_{i}e^{i\vec{q}\cdot\vec{R}_i}\:D^{(1)}_{\vec{q}}
\cdot\left(\vec{S}_i\cdot\boldsymbol{\hat\sigma}\right)_{\sigma^{\prime}\sigma}.
\end{multline}
Here we have defined
\begin{equation}\label{sn:eq:D1}
  D^{(1)}_{\vec{q}}=-\frac{1}{\pi}\int\limits^{+\infty}_{-\infty}dE\:f_-(E)
 \Im \frac{1}{N}\sum_{\vec{k}}G^{(0)}_{\vec{k}}(E-\mu)G^{(0)}_{\vec{k}+\vec{q}}(E-\mu).
\end{equation}
Because of 
\begin{equation}\label{sn:eq:Ssigma}
  \vec{S}_i\cdot\boldsymbol{\hat\sigma}=\frac{1}{2}\begin{pmatrix}
      S^z_i&S^-_i\\ 
      S^+_i&-S^z_i\end{pmatrix}
\end{equation}
one realizes that the first term in Eq.~(\ref{sn:eq:averc+c1stOrder}) does
not contribute to Eq.~(\ref{sn:eq:map}), while the second term gives rise to
an effective Hamiltonian like Eq.~(\ref{sn:eq:Hf}) with the following
effective exchange integrals:
\begin{equation}\label{sn:eq:J1ij}
  \hat J^{(1)}_{ij}=\frac{1}{N}\sum_{\vec{q}}\hat J^{(1)}(\vec{q})
\:e^{-i\vec{q}\cdot(\vec{R}_i-\vec{R}_j)},
\end{equation}
\begin{eqnarray}\label{sn:eq:J1q}
  \hat J^{(1)}(\vec{q})&=&-\frac{1}{2}J^2\:D^{(1)}_{\vec{q}}\\\nonumber
  &=&-\frac{1}{2}J^2\hbar^2\sum_{\vec{k}}
\frac{f_-(\varepsilon(\vec{k}+\vec{q}))-f_-(\varepsilon(\vec{k}))}
{\varepsilon(\vec{k}+\vec{q})-\varepsilon(\vec{k})}.
\end{eqnarray}
This is the classical RKKY exchange, which comes out in first order of
our Green function procedure.

It is well-known that Eq.~(\ref{sn:eq:J1q}) can be equivalently derived by
conventional second-order perturbation theory starting from an
unpolarized conduction electron gas. In order to incorporate the
exchange-induced spin polarization of the conduction electrons
to higher order, it almost suggests itself to modify approximation
(\ref{sn:eq:1stOrder}) by replacing in the exact expression
[Eq.~(\ref{sn:eq:combined})] the restricted Green function by the full, all
polarization processes containing single-electron Green function
[Eq.~(\ref{sn:eq:SingleGreen})]: 
\begin{eqnarray}\label{sn:eq:reG}
  \hat
  G^{\sigma^{\prime\prime}\sigma}_{\vec{k}^{\prime},\vec{k}+\vec{q}}(E)
\longrightarrow&&\delta_{\sigma^{\prime\prime}\sigma}
 \delta_{\vec{k}^{\prime}\vec{k}+\vec{q}}G_{\vec{k}+\vec{q}\sigma}(E),\\
  \hat
  G^{\sigma^{\prime}\sigma^{\prime\prime}}_{\vec{k},\vec{k}^{\prime}}(E)
\longrightarrow&&\delta_{\sigma^{\prime}\sigma^{\prime\prime}}
 \delta_{\vec{k}\vec{k}^{\prime}}G_{\vec{k}\sigma^{\prime}}(E).
\end{eqnarray}
For the required expectation value [Eq.~(\ref{sn:eq:averc+cs})], we obtain an
expression similar to Eq.~(\ref{sn:eq:averc+c1stOrder}), only
$D^{(1)}_{\vec{q}}$ has to be replaced by 
\begin{eqnarray}\label{sn:eq:Dq}
\lefteqn{  D^{\sigma\sigma^{\prime}}_{\vec{q}}=-\frac{1}{\pi}
\int\limits^{+\infty}_{-\infty}dE\:f_-(E)\Im \frac{1}{N}\sum_{\vec{k}}\times}\\\nonumber
&&\hspace{5em}\times\left(G^{(0)}_{\vec{k}}(E-\mu)
\:G_{\vec{k}+\vec{q}\sigma}(E-\mu)+\right.\\\nonumber
&&\hspace{7em}\left.+G^{(0)}_{\vec{k}+\vec{q}}(E-\mu)
\:G_{\vec{k}\sigma^{\prime}}(E-\mu) \right).\\\nonumber
\end{eqnarray}
Exploiting the inversion symmetry, one proves
\begin{equation}\label{sn:eq:Dsymm}
  D^{\uparrow\downarrow}_{\vec{q}}= D^{\downarrow\uparrow}_{\vec{q}}
=\frac{1}{2}\sum_{\sigma}D^{\sigma\sigma}_{\vec{q}}.
\end{equation} 
This guarantees an isotropic effective exchange operator
Eq.~(\ref{sn:eq:Hf}) with the exchange-integral
\begin{multline}\label{sn:eq:Jq}
  \hat
  J(\vec{q})=-\frac{J^2}{4}\int\limits^{+\infty}_{-\infty}dE\:f_-(E)\frac{1}{N}
\!\sum_{\vec{k}\sigma}\bigg(-\frac{1}{\pi}\Im\Big(G^{(0)}_{\vec{k}}(E-\mu)\times\Big.\bigg.\\
\bigg.\Big.\:\times G_{\vec{k}+\vec{q}\sigma}(E-\mu)\Big)\bigg).
\end{multline}
Via $G_{\vec{k}+\vec{q}\sigma}(E-\mu)$ the effective exchange is a
functional of the electronic self-energy Eq.~(\ref{sn:eq:MCDA}), and
therefore receives a distinct temperature and electron density dependence.

To obtain the magnetic properties of the local-moment system from the
effective operator [Eq.~(\ref{sn:eq:Hf})], we use the spin Green function 
\begin{equation}\label{sn:eq:MagnonP}
  P^{(a)}_{ij}(E)=\bigl\langle\!\bigl\langle S^+_i;e^{aS^z_j}S^-_j\bigr\rangle\!\bigr\rangle_{E}
\end{equation}
first proposed by Callen\cite{CAL63}. $a$ is a real number, which
eventually helps to derive several spin-correlation functions by
applying the spectral theorem to $P^{(a)}_{ij}(E)$ and a proper
differentiation with respect to $a$. The equation of motion of
$P^{(a)}_{ij}(E)$ reads
\begin{multline}\label{sn:eq:eqmPa}
  E P^{(a)}_{ij}(E)=2\hbar\delta_{ij}\langle
  A_a\rangle-2\hbar\sum_{m}\hat J_{im}\times\\
\times \bigl\langle\!\bigl\langle\left(S^+_m S^z_i-S^+_i S^z_m
\right);e^{aS^z_j}S^-_j\bigr\rangle\!\bigr\rangle_E.
\end{multline}
Here we have abbreviated
\begin{equation}\label{sn:eq:Aa}
  \langle A_a\rangle=\left\langle[S^+_i,e^{aS^z_i}S^-_i]_-\right\rangle.
\end{equation}
It is well-known that a simple random-phase-approximation decoupling 
of the higher Green
function on the right-hand side of Eq.~(\ref{sn:eq:eqmPa}) yields
surprisingly convincing results in the low- and high- temperature 
region (''Tyablikov-approximation''). Doing so, after a
Fourier-transformation we obtain
\begin{equation}\label{sn:eq:RPA_P}
  P^{(a)}_{\vec{q}}(E)=\frac{2\hbar^2\langle A_a\rangle}{E-E(\vec{q})+i0^+}.
\end{equation}
In this approximation the magnon energies $E(\vec{q})$ are real:
\begin{equation}\label{sn:eq:MagnonEnergies}
  E(\vec{q})=2\hbar\langle S^z\rangle\bigl(\hat J_{\vec{0}}-\hat J(\vec{q})\bigr),
\end{equation}
\begin{equation}\label{sn:eq:J0}
  \hat J_0=\hat J(\vec{q}=\vec{0})=\sum_{i}\hat J_{ij}.
\end{equation}
Following the Callen-method\cite{CAL63} for the local-moment
magnetization one obtains
\begin{equation}\label{sn:eq:Sz}
  \langle
  S^z\rangle=\hbar\frac{(1+S+\varphi)\:\varphi^{2S+1}+(S+\varphi)\:(1+\varphi)^{2S+1}}
 {(1+\varphi)^{2S+1}-\varphi^{2S+1}}.
\end{equation}
$\varphi$ is the average magnon number,
\begin{equation}\label{sn:eq:magnumber}
\varphi(S)=\frac{1}{N}\sum_{\vec{q}}\frac{1}{e^{\beta E(\vec{q})}-1},
\end{equation}
which determines $\langle S^z\rangle$ and many other spin
correlations.
Some typical examples are
\begin{eqnarray}\label{sn:eq:SmSp}
\langle S^-S^+\rangle&=&2\hbar\langle S^z\rangle\:\varphi(S),\\
\langle (S^z)^3\rangle&=&\hbar^3S(S+1)\varphi(S)+\hbar^2\langle
S^z\rangle\bigl(S(S+1)+\bigr.\nonumber\\
\label{sn:eq:Sz3}
&&\bigl.+\varphi(S)\bigr)-\hbar\langle (S^z)^2\rangle\bigl(1+3\varphi(S)\bigr).
\end{eqnarray}
Via Eq.~(\ref{sn:eq:Jq}) all spin correlations
[Eqs.~(\ref{sn:eq:Sz})-(\ref{sn:eq:Sz3})]
are fixed by the electron self-energy [Eq.~(\ref{sn:eq:MCDA})].

The key quantity of ferromagnetism is the Curie temperature
$T_{\textrm{C}}$. Performing the limiting process 
\begin{equation}\label{sn:eq:limTc}
T\longrightarrow T^{(-)}_{\textrm{C}}; \langle S^z\rangle\longrightarrow
0^+
\end{equation} 
one finds, from Eq.~(\ref{sn:eq:Sz}),
\begin{equation}\label{sn:eq:Tc}
  k_{\textrm{B}}T_{\textrm{C}}=\frac{2}{3}\hbar^2S(S+1)
\left[ \frac{1}{N}\sum_{\vec{q}}\left(\hat J_{\vec{0}}
-\hat J(\vec{q})\right)^{-1}_{T_{\textrm{C}}}\right]^{-1}.
\end{equation}
The effective exchange integrals are temperature-dependent and have to
be used here for $T\rightarrow T_{\textrm{C}}$.

For the results presented in Sec.~\ref{sn:sc:MagProp}, we have used
the self-energy approach of Ref.~\onlinecite{NRM97}. The same approach
has been applied in Refs.~\onlinecite{SNO01a, SNO01b,SMN01, REN99, MSN01},
so that we can refrain here from presenting details of the method. As
explained after Eq.~(\ref{sn:eq:MCDA}) the self-energy of
Ref.~\onlinecite{NRM97} depends on three types of expectation values:
purely electronic terms, which are accessible with the single-particle
Green function [Eq.~(\ref{sn:eq:SingleGreen})]; pure local-spin correlation
such as Eqs.~(\ref{sn:eq:Sz}), (\ref{sn:eq:SmSp}), (\ref{sn:eq:Sz3}); and
mixed itinerant-electron-local-moment correlations, which are
expressible by the ''higher'' Green functions [Eqs.~(\ref{sn:eq:spinflipfunc})
and (\ref{sn:eq:isingfunc})]. Eventually, we have a closed system of
equations that can be solved self-consistently for the desired
electronic and magnetic properties of the KLM. Some typical results are
discussed in Sec.~\ref{sn:sc:MagProp}.
\section{Magnetic Properties}\label{sn:sc:MagProp}
We have evaluated our theory for a s.c. lattice (bandwidth: $W=1$ eV) to
find out the magnetic properties of the ferromagnetic KLM.
The latter are, however, strongly correlated with the electronic
properties, so that they can not be understood without referring to the
itinerant-electron subsystem. 
\begin{figure}[h]
\includegraphics[width=0.4\textwidth]{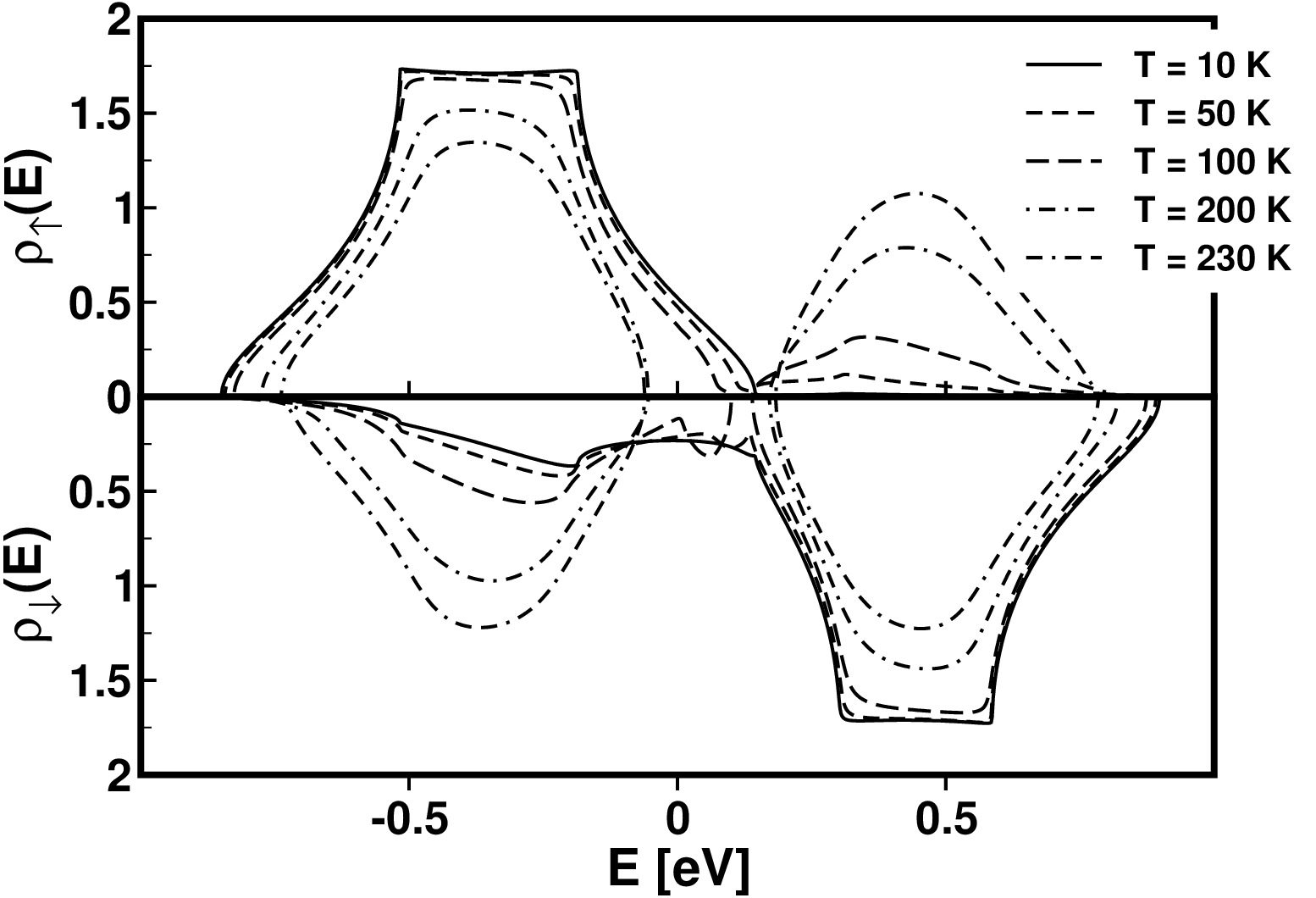}
\caption{Temperature-dependent quasiparticle density of states $\rho_{\sigma}(E)$
        as function of energy (Kondo-lattice model).
        Parameters: $W=1$ eV, $S=7/2$, $J=0.2$ eV, $n=0.2$.
       For these parameters the theory yields: $T_{\textrm{C}}=232$ K.
       The upper half for $\sigma=\uparrow$, the lower half for $\sigma=\downarrow$.}
\label{sn:fig:QDOS}
\end{figure}
Fig.~\ref{sn:fig:QDOS} shows the quasiparticle density of states (QDOS)
$\rho_{\sigma}(E)$ for a band occupation $n=0.2$ and a moderate
exchange coupling $J=0.2$ eV. The self-consistently determined Curie
temperature turns out to be $T_{\textrm{C}}=232K$. In the low-temperature,
ferromagnetic phase the QDOS exhibits a remarkable
temperature-dependence. For $T\rightarrow 0$ the $\uparrow$ spectrum
becomes rather simple (solid line in the upper half of Fig.~\ref{sn:fig:QDOS})
since the local-moment system is ferromagnetically saturated. A
$\uparrow$ electron has therefore no chance to exchange its spin with
the localized spins. That means that only the Ising-like interaction
term in Eq.~(\ref{sn:eq:Hsf2quant}) really works, leading to a rigid shift
of about $-\frac{1}{2}JS$ of $\rho_{\uparrow}(E)$ with respect to the
''free'' Bloch DOS $\rho_{0}(E)$. The down-spin QDOS, however, is much
more complicated, because a $\downarrow$ electron can exchange its spin
with the saturated localized spins. That can, e.g., be done by emitting a
magnon $\hbar\omega(\vec{q})$, where the electron reverses its spin from
$\downarrow$ to $\uparrow$. Consequently, this part of the spectrum
(''scattering states'') exactly coincides with
$\rho_{\uparrow}(E)$. However, the
$\downarrow$ electron has another possibility to exchange its spin with
the perfectly aligned localized spins. It can polarize its local
spin surrounding by repeated magnon emission and reabsorption. This can
even lead to a bound state, which we call the ''magnetic polaron''. For
the special case of a single $\downarrow$ electron in an otherwise empty
conduction band coupled to a ferromagnetically saturated spin system the
formation of the magnetic polaron can rigorously be shown
\cite{SMA81,NDM85} (see Fig.~1 in Ref.~\onlinecite{MSN01}). Formation of
a polaron costs more energy than magnon emission. Polaron states
therefore build the upper part of the $\downarrow$ spectrum.

Magnon emission by the $\downarrow$ electron is equivalent to magnon
absorption of a $\uparrow$ electron, with one exception: magnon
absorption can occur only if there are magnons in the system. This is not
the case at $T=0$ in the ferromagnetic saturation. At finite
temperature, however, there are magnons to be absorbed 
by $\uparrow$ electrons. Consequently, scattering
states appear in the $\uparrow$ spectrum too. Because of the possible
spin exchange both spin spectra occupy exactly the same
energy region at all $T > 0$. It is interesting to observe that even for such a
moderate coupling $J=0.2$ eV ($W=1$ eV) there opens a gap between a lower
and an upper quasiparticle subband. In the lower subband the electron
hops mainly over lattice sites, where it orients its spin parallel to the
local moment, either without or with preceding spin-flip by
magnon emission (absorption). The upper subband consists of polaron
states with finite lifetimes. Note that such elementary processes are
not recognizable when for mathematical simplicity classical spins
($S\rightarrow\infty$) are assumed \cite{FUR99,MLS95,HVO00}. The QDOS
in Fig.~\ref{sn:fig:QDOS} clearly demonstrates that the conduction
electrons are not at all fully spin polarized as is often used in
DMFT treatments of the KLM\cite{FUR96,HVO00}. That is true only in the
unphysical limit $S\rightarrow\infty.$

Let us now concentrate the rest of the discussion on the magnetic
properties of the KLM. They are determined by the respective behaviour
of the effective exchange integrals $\hat J_{ij}$.

\begin{figure}[ht]
\includegraphics[width=0.4\textwidth]{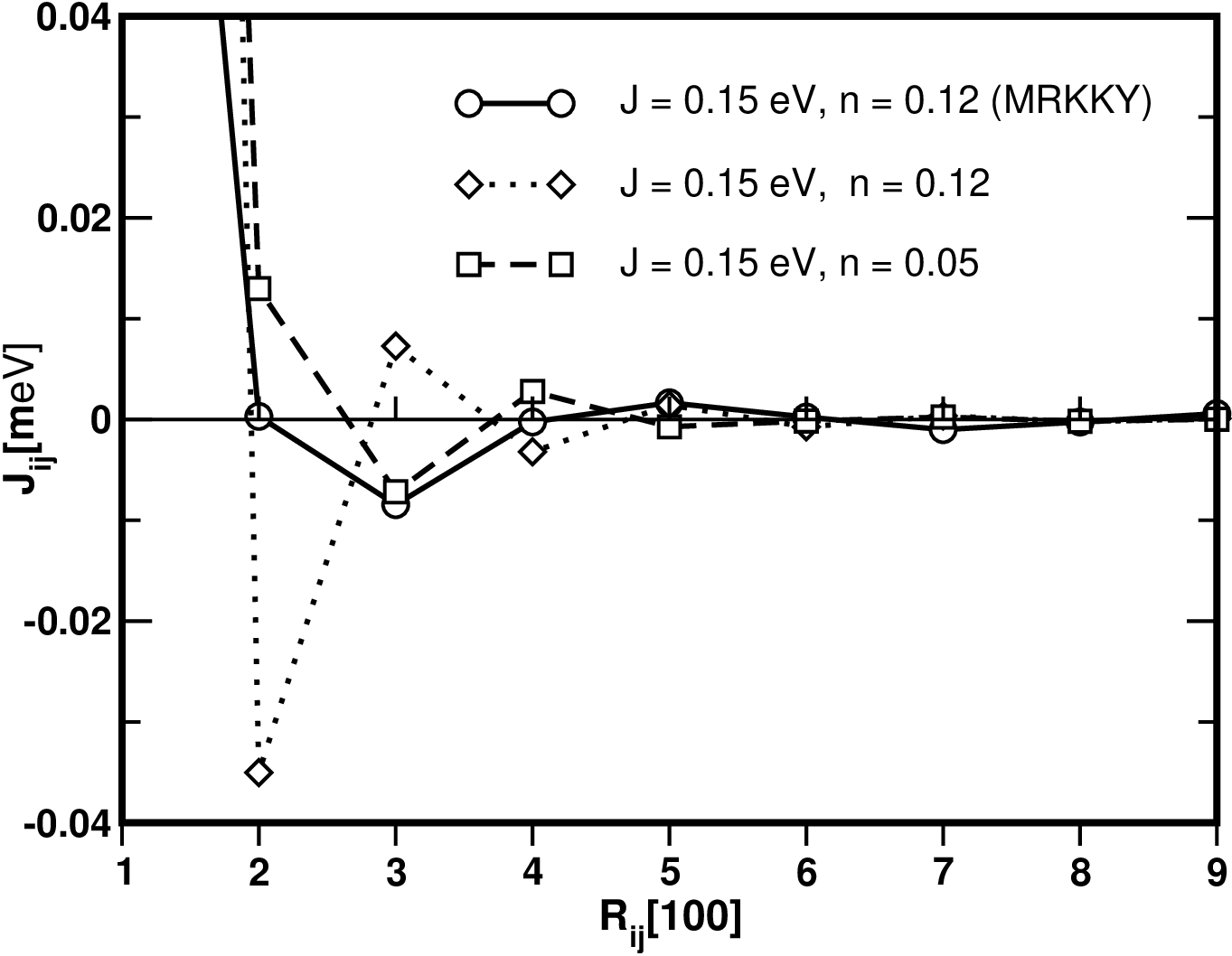}
\caption{Conventional RKKY exchange integrals $\hat J_{ij}$ as 
        a function of the distance $R_{ij}$ between lattice sites 
        in (100) direction for various band occupation $n$ 
        and different interband exchange couplings $J$. 
        Parameters: $T=0K$, $W=1$ eV, s.c. lattice. For comparison
                a respective example of the modified RKKY model is included.}
\label{sn:fig:JijConv}
\end{figure}

Fig.~\ref{sn:fig:JijConv} shows the distance dependence of $\hat J_{ij}$
as it follows from the conventional RKKY [Eq.~(\ref{sn:eq:J1ij})]. We recognize
the well-known oscillating and long-range behavior. According to
Eq.~(\ref{sn:eq:J1q}) two parameters influence the oscillation: the
interband exchange $J$ and the band occupation $n$. The ''amplitude'' of
the oscillation is proportional to $J^2$, while the ''period'' is fixed
via the chemical potential $\mu$ by the conduction electron density
$n$. A remarkable temperature dependence is not observed.

\begin{figure}[h]
\begin{center}
\includegraphics[width=0.4\textwidth]{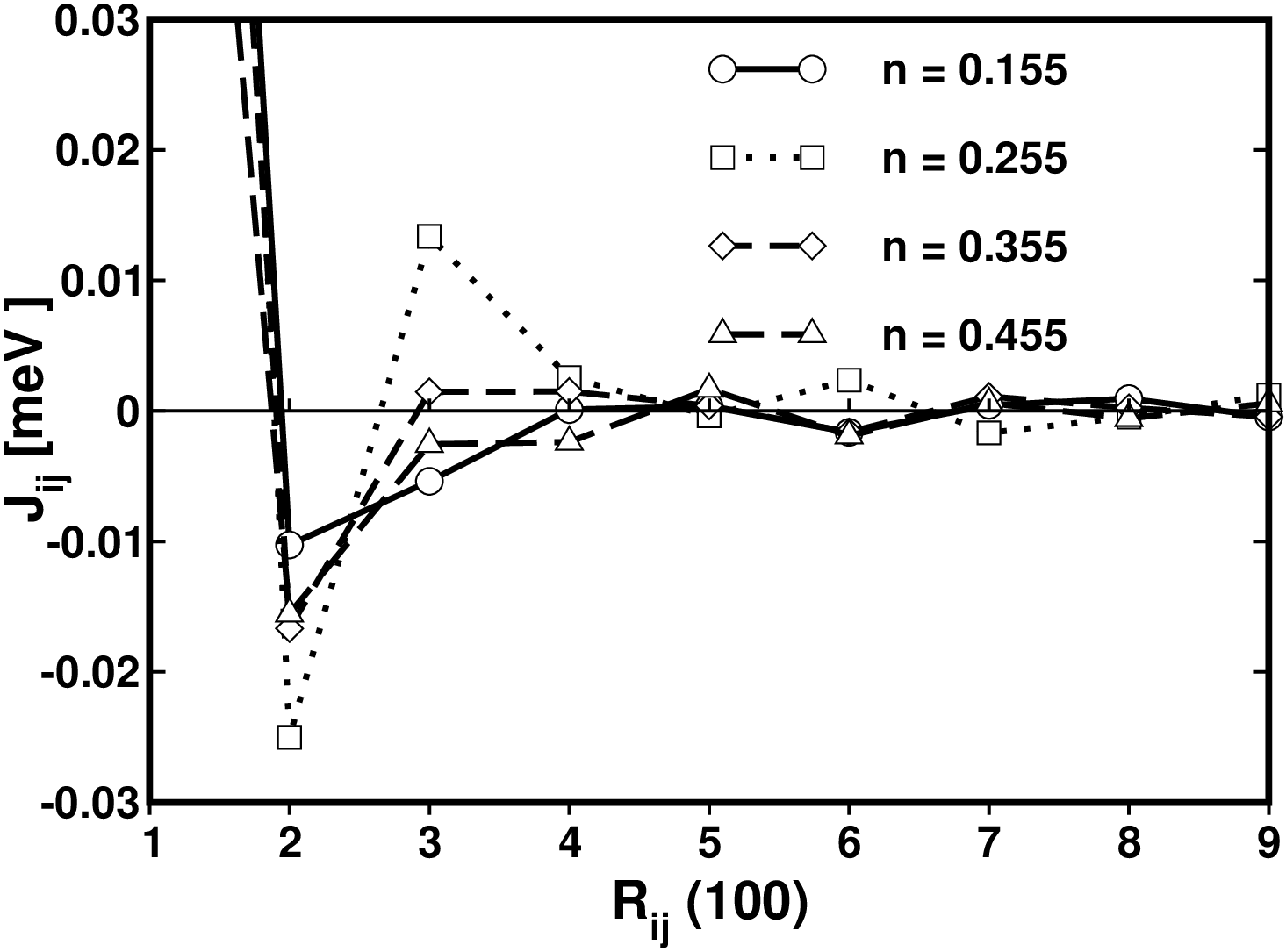}
\end{center}
\caption{Effective exchange integrals $\hat J_{ij}$ of the ''modified''
         RKKY model as a function of the distance $R_{ij}$ in (100) direction
         for various band occupations. Parameters: 
         $T=0K$, $J=0.1$ eV, $W=1$ eV, s.c.~lattice, $S=7/2$.}
\label{sn:fig:Jija}
\end{figure}

\begin{figure}[ht]
\begin{center}
\includegraphics[width=0.4\textwidth]{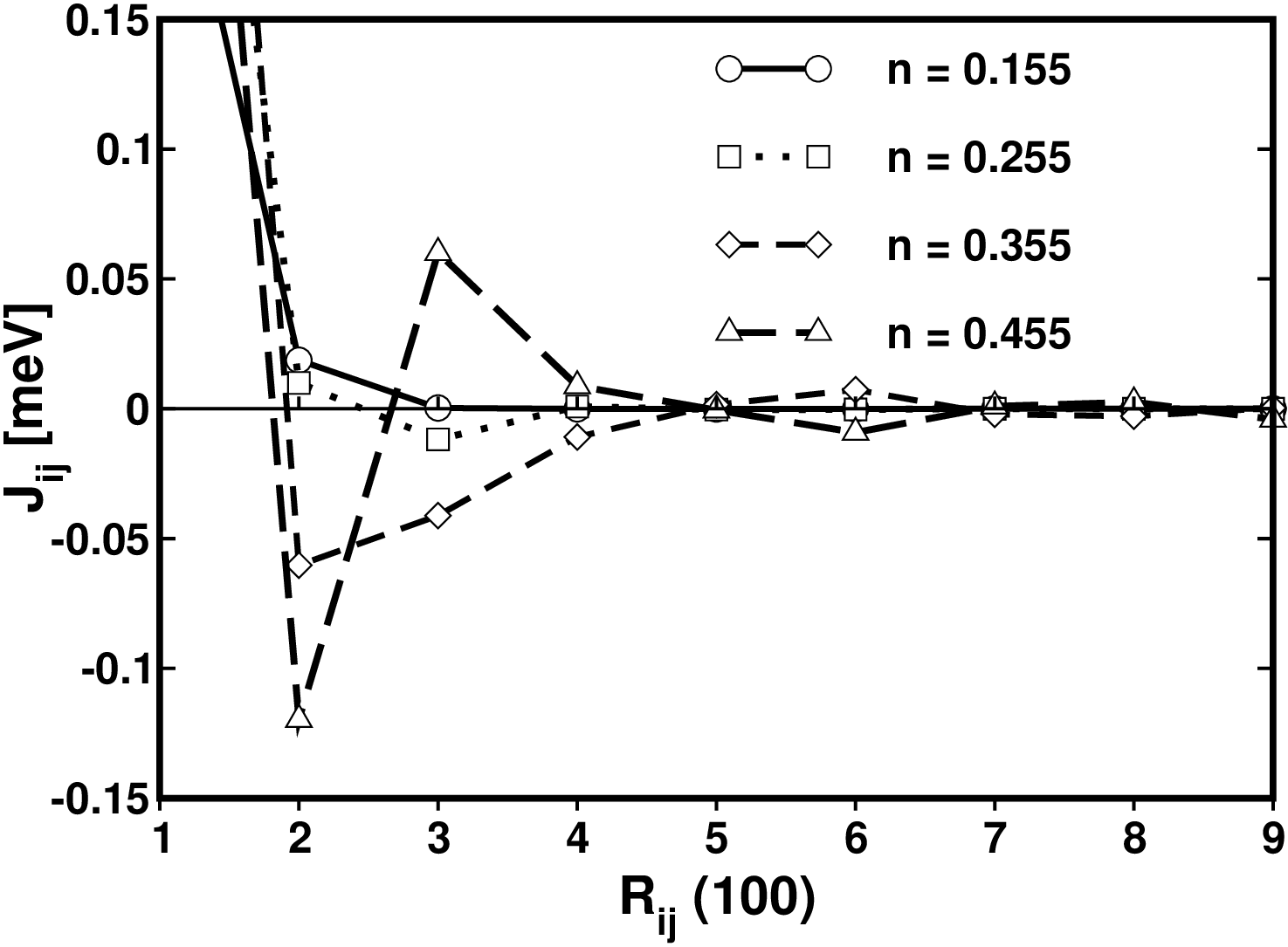}
\end{center}
\caption{The same as in Fig.~\ref{sn:fig:Jija}, but for $J=0.2$ eV.}
\label{sn:fig:Jijb}
\end{figure}

The oscillations of the effective exchange integrals $\hat J_{ij}$ are
less regular when higher-order polarization effects in the conduction
band are taken into account [Eq.~(\ref{sn:eq:Jq})]. Fig.~\ref{sn:fig:Jija} and
Fig.~\ref{sn:fig:Jijb} show two examples for two different $J$'s and several
band occupations $n$.
The nonregular $J$ dependence results from the
fact, that $J$ appears in Eq.~(\ref{sn:eq:Jq}) not only as a prefactor $J^2$
but also in a complicated manner via the Green function
$G_{\vec{k}+\vec{q}\sigma}(E-\mu)$ of the polarized itinerant electron
system.  
Of similar complexity is the $n$ dependence. In the conventional RKKY
[Eq.~(\ref{sn:eq:J1q})] it comes into play only through the chemical potential
in the Fermi function, while in the full expression [Eq.~(\ref{sn:eq:Jq})], 
several correlation functions as $\langle n_{\sigma}\rangle$,
$\langle S^zn_{\sigma}\rangle$, $\langle
S^{-\sigma}c^{\dagger}_{\sigma}c^{}_{-\sigma}\rangle$,\ldots 
[Eqs.~(\ref{sn:eq:occnumber}), (\ref{sn:eq:gamma}), (\ref{sn:eq:delta})]
contribute to the $n$ dependence.

\begin{figure}[ht]
\includegraphics[width=0.4\textwidth]{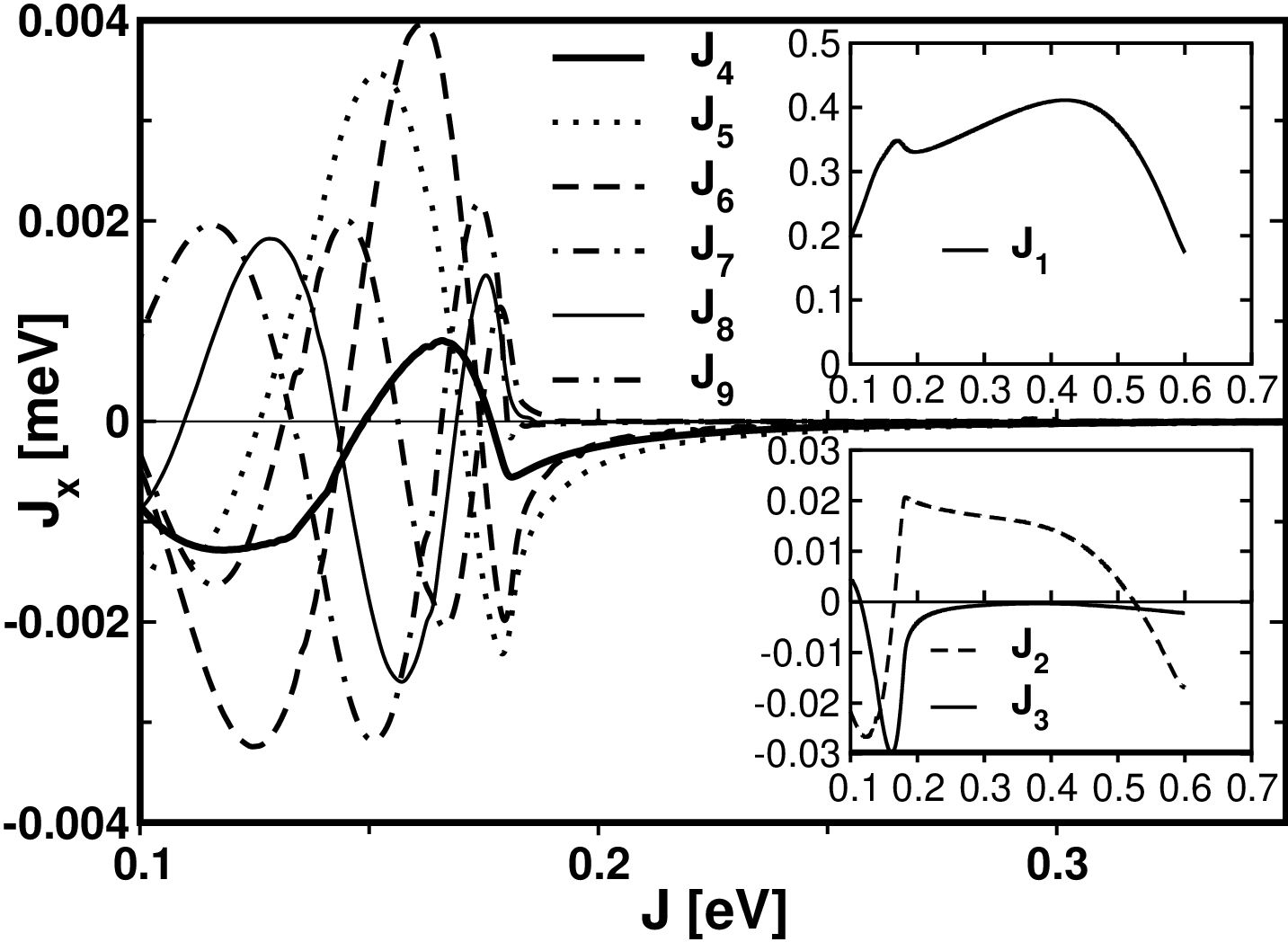}
\caption{Effective exchange integrals $J_x$ between $x$th
         nearest neighbors ($x=1, 2,\ldots,9$) in (100) direction
         (s.c.~lattice) as a function of the interband exchange 
         coupling $J$. Parameters: $T=0$ K, $n=0.2$, $W=1$ eV,
         $S=7/2$. Note the different energy scales.}
\label{sn:fig:Jij_J}
\end{figure}

To demonstrate the influence of the interband exchange
coupling $J$ on the effective Heisenberg-exchange integrals 
$\hat J_{ij}$ in more detail, in Fig.~\ref{sn:fig:Jij_J} we plot the $J$
dependence
of $J_x$, $x=1,\ldots,9$. $J_x$ is the effective exchange integral
between sites separated along the (100) direction by $x$ lattice  
constants.\footnote{Note that $J_2$ is already the
fourth-nearest neighbour.}.
$J_4,\ldots,J_9$ show an oscillatory structure in the weak-coupling
region, being, one order of magnitude smaller than $J_{2,3}$ and
even two orders of magnitude smaller than $J_1$. For $J>0.2$ eV
$J_3,\ldots,J_9$ are so strongly damped that the magnetism of the KLM can
be described by a Heisenberg model with next- and next-nearest-neighbor
exchanges only. In the weak-coupling regime, however, the higher terms 
$J_3,\ldots,J_9$ can not be neglected. Similar statements can be found
in\cite{FUR96}. Maybe this is an explanation why CMR materials with high
$T_{\textrm{C}}$ can be described reasonably well\cite{PAH96} by a
short-range Heisenberg model, while other materials with lower
$T_{\textrm{C}}$ exhibit strong deviations\cite{HDC98,FDH98,DHF00}. In
any case the main contribution to the exchange coupling stems from the
next-nearest-neighbors. $J_1$ determines the stable magnetic configuration. The
negative slope of $J_1$ for $J>0.5$ eV indicates already a breakdown of
the ferromagnetic order, which is more carefully inspected below.

\begin{figure}[h]
\includegraphics[width=0.4\textwidth]{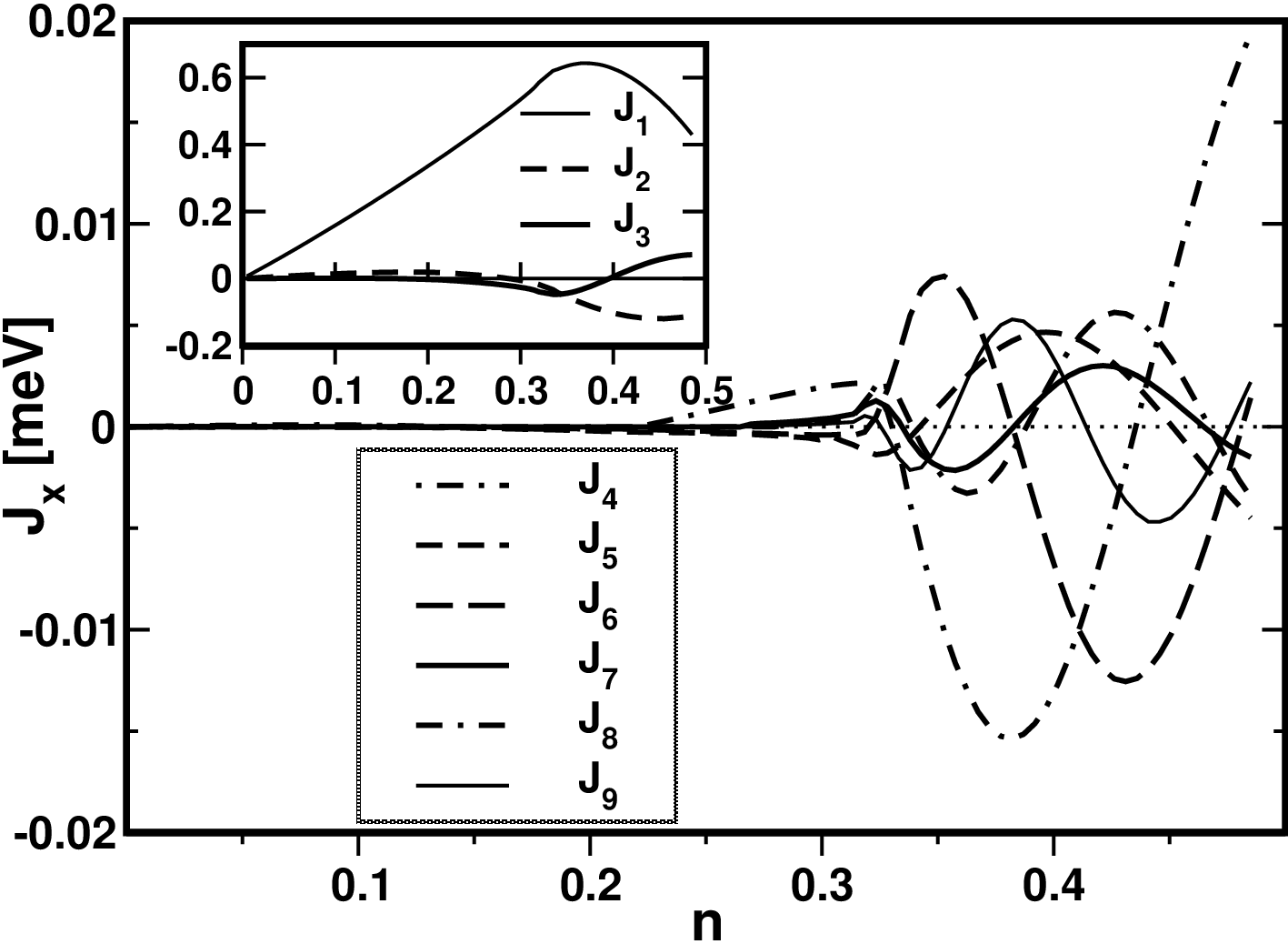}
\caption{The same effective exchange integrals as in
  Fig.~\ref{sn:fig:Jij_J}, but now as a function of the band occupation
 $n$ for $J=0.2$ eV.}
\label{sn:fig:Jij_n}
\end{figure}

The effective exchange integrals $J_x$ exhibit an interesting 
electron-density dependence (Fig.~\ref{sn:fig:Jij_n}). For low densities up to
$n=0.3$ only $J_1$ is of importance. A Heisenberg model with
nearest-neighbour 
exchange will appropriately describe the magnetic properties
of the KLM. 
For larger $n$, however, the exchange interaction becomes very long-range, even
$J_9$ is not negligible. From Fig.~\ref{sn:fig:Jij_J} and
\ref{sn:fig:Jij_n} we learn that the interplay between local interband
exchange $J$ and band occupation $n$ does lead to a rather complicated
behaviour of the effective RKKY exchange integrals.

\begin{figure}[h]
\includegraphics[width=0.4\textwidth]{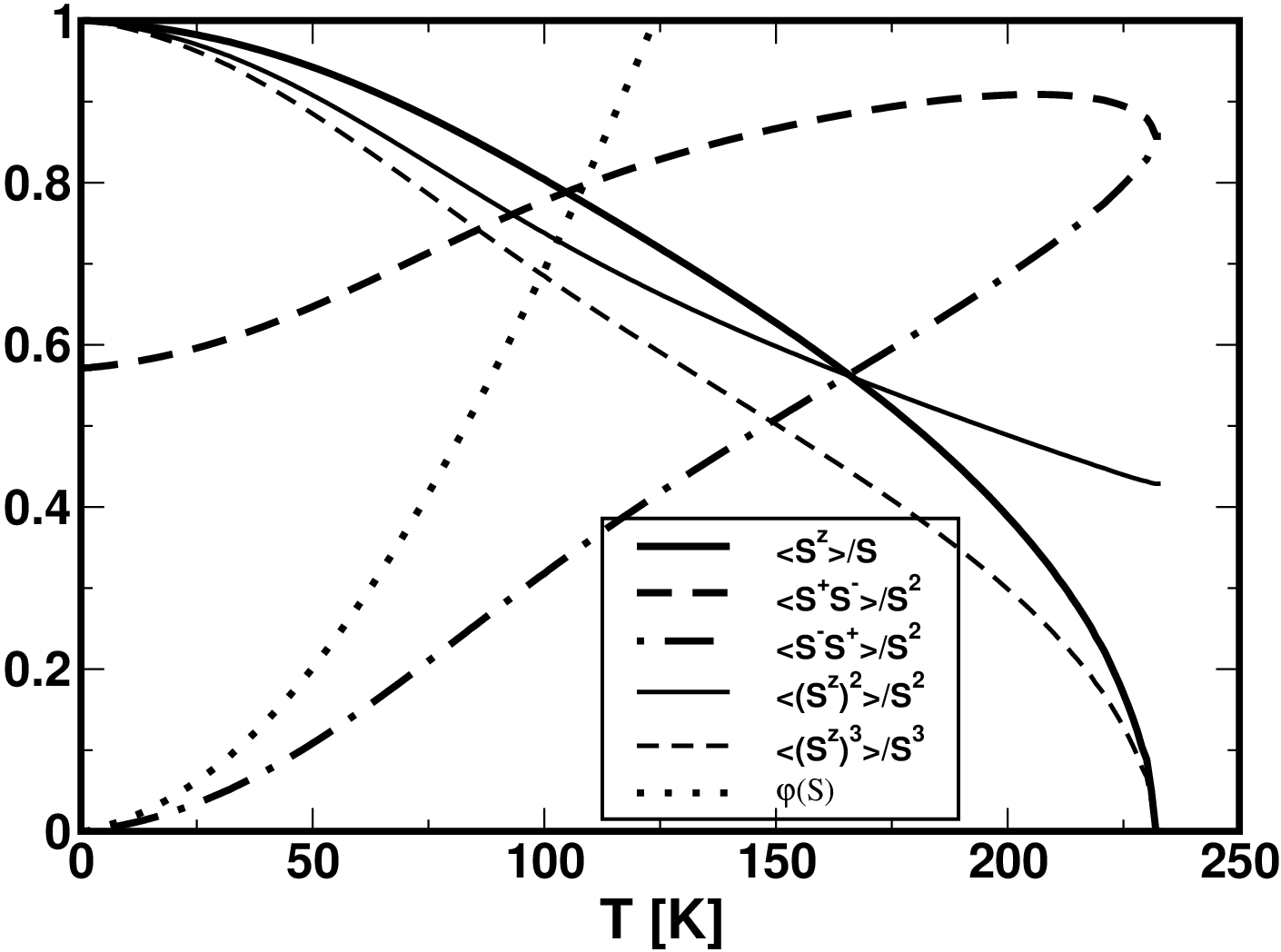}
\caption{Several spin correlation functions of the local-moment system
         in dependence on the temperature $T$. Parameters: $J=0.2$ eV,
         $n=0.2$, $S=7/2$, s.c.~lattice, $W=1$ eV. The theory, presented
         in the text, self-consistently yields $T_{\textrm{C}}=232$ K.}
\label{sn:fig:spincT}
\end{figure}

The effective exchange integrals determine the magnetic properties of
the KLM. On the other hand, they depend on the electron self-energy which
is strongly influenced by magnetic correlation functions. Typical
examples of the latter are plotted in Fig.~\ref{sn:fig:spincT},
calculated for $n=0.2$ and $J=0.2$ eV.
The electronic part of the self-consistent solution, represented by the
temperature dependent QDOS, is shown in Fig.~\ref{sn:fig:QDOS}. The
Curie temperature is found as $T_{\textrm{C}}=232$ K. The local-moment
magnetization $\langle S^z\rangle$ is of Brillouin function type, and the
qualitative shapes of the other spin correlations are also familiar from the pure
Heisenberg model.

\begin{figure}[h]
\includegraphics[width=0.4\textwidth]{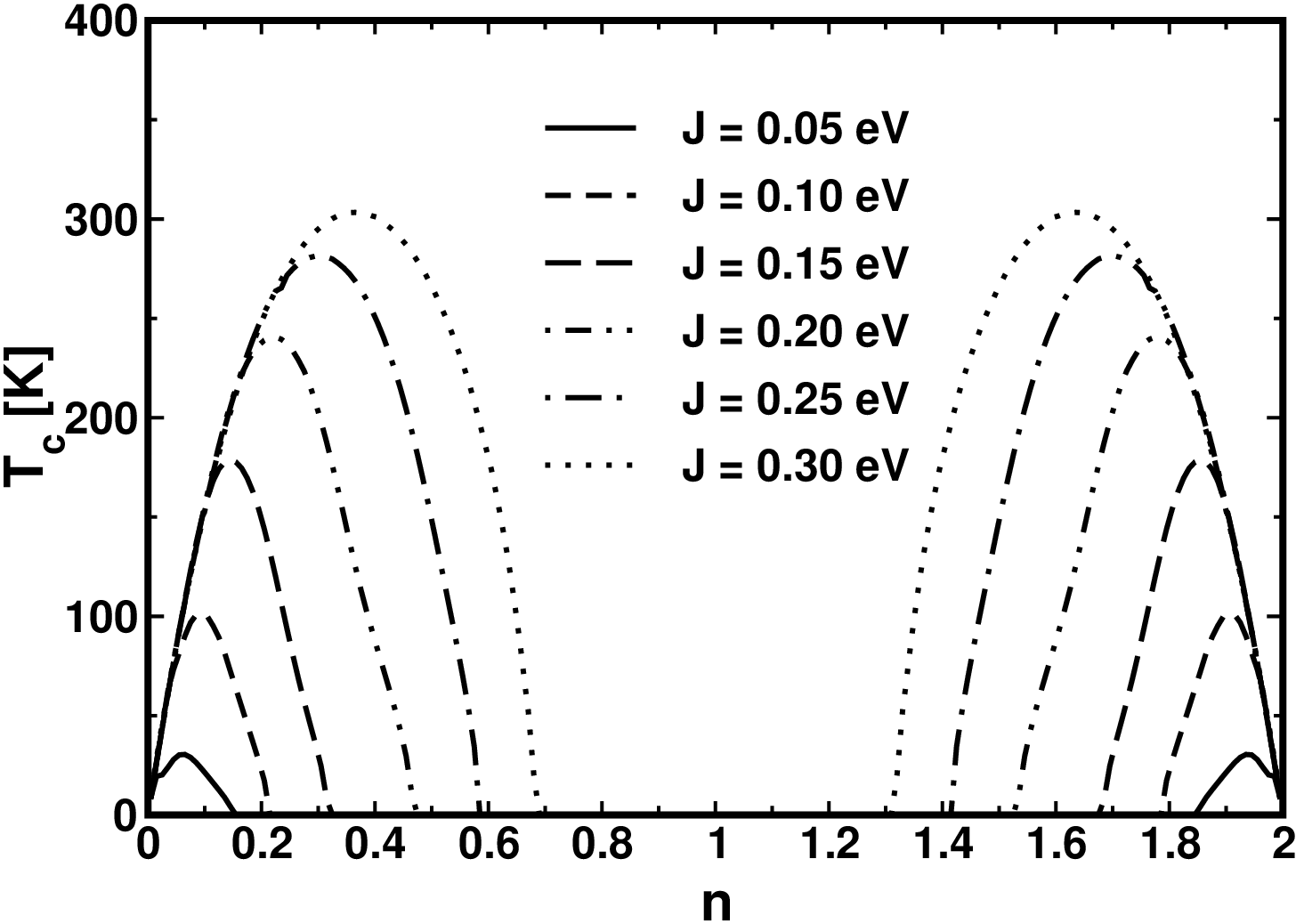}
\caption{Curie temperature as a function of the band occupation $n$
         for various interband exchange couplings 
         ($S=7/2$, s.c.~lattice, $W=1$ eV), calculated by use of the 
         ''modified'' RKKY. model.}
\label{sn:fig:Tc_n}
\end{figure}

The key-quantity of ferromagnetism is the Curie temperature
$T_{\textrm{C}}$. $T_{\textrm{C}}>0$ comes out as a consequence of an
indirect coupling between the local moments, mediated by spin
polarization of itinerant electrons. Therefore, a strong particle
density dependence of $T_{\textrm{C}}$ has to be
expected. Figure \ref{sn:fig:Tc_n} shows $T_{\textrm{C}}(n)$ for various
$J$ from the weak to moderate coupling regime, calculated according to
[Eq.~(\ref{sn:eq:Tc})].
Ferromagnetism appears for low particle (hole) densities, where the
ferromagnetic region increases with increasing $J$. The
$T_{\textrm{C}}$ values are of realistic order. A similar $n$ dependence
of $T_{\textrm{C}}$ has been found in Ref.~\onlinecite{CMI01} within an
extended multi-band KLM. No ferromagnetism appears around half-filling
($n=1$). It can be speculated that antiferromagnetism, which is not
considered within our approach, becomes stable near $n=1$. It is
\begin{figure}[h]
\includegraphics[width=0.375\textwidth]{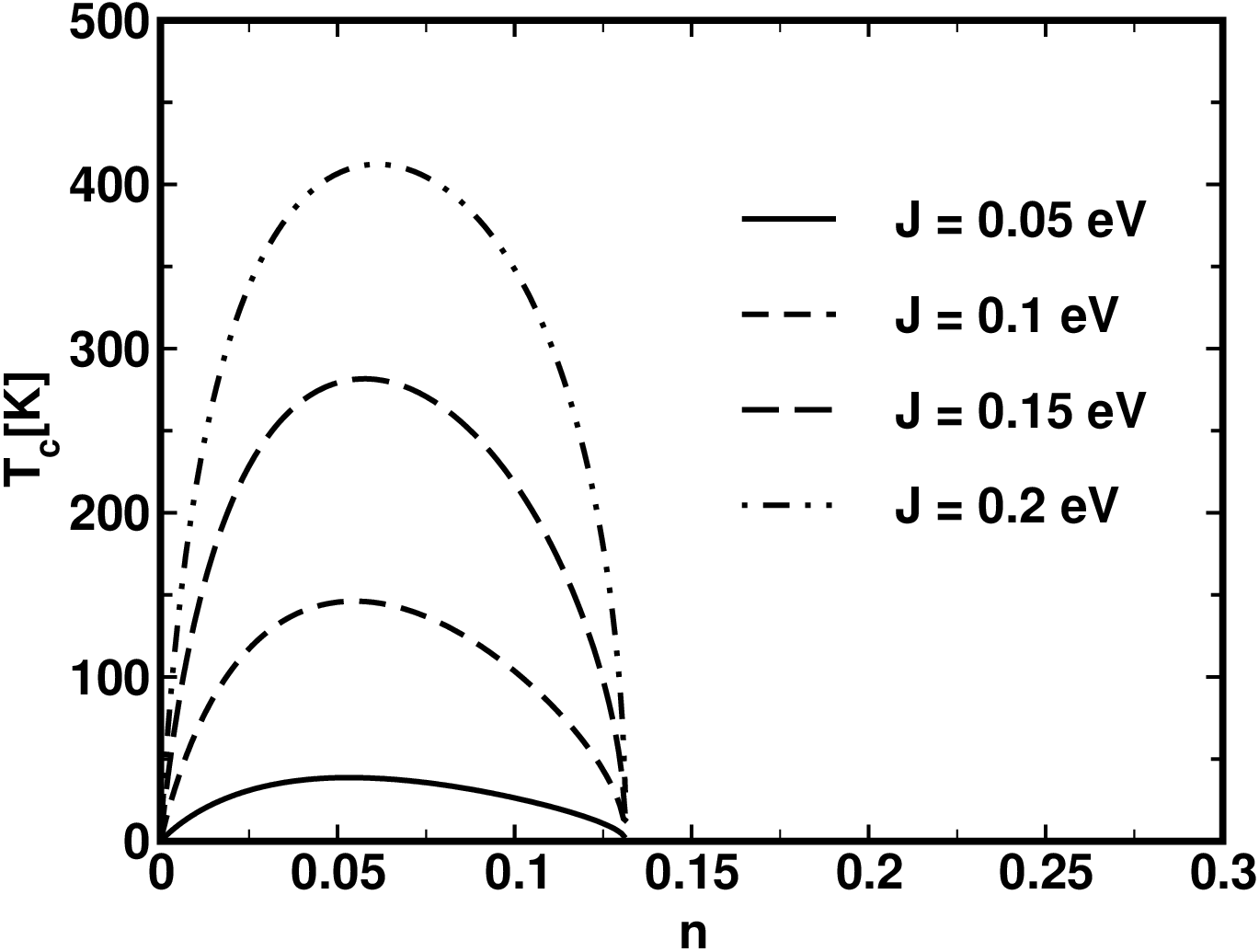}
\caption{The same as in Fig.~\ref{sn:fig:Tc_n}, but calculated by use of
         the ''conventional'' RKKY model.}
\label{sn:fig:Tc_n_Conv}
\end{figure}
interesting to compare the results of Fig.~\ref{sn:fig:Tc_n} with their
conterparts from conventional RKKY (Fig.~\ref{sn:fig:Tc_n_Conv}), using
the same model parameters and the same $T_{\textrm{C}}$-formula
[Eq.~(\ref{sn:eq:Tc})]. The $T_{\textrm{C}}$ values are higher, but the region
of $n$ where ferromagnetism appears, is distinctly narrower. The
critical $n$, where $T_{\textrm{C}}$ vanishes again, is independent of
$J$, in contrast to the full theory in Fig.~\ref{sn:fig:Tc_n}.

\begin{figure}[h]
\includegraphics[width=0.4\textwidth]{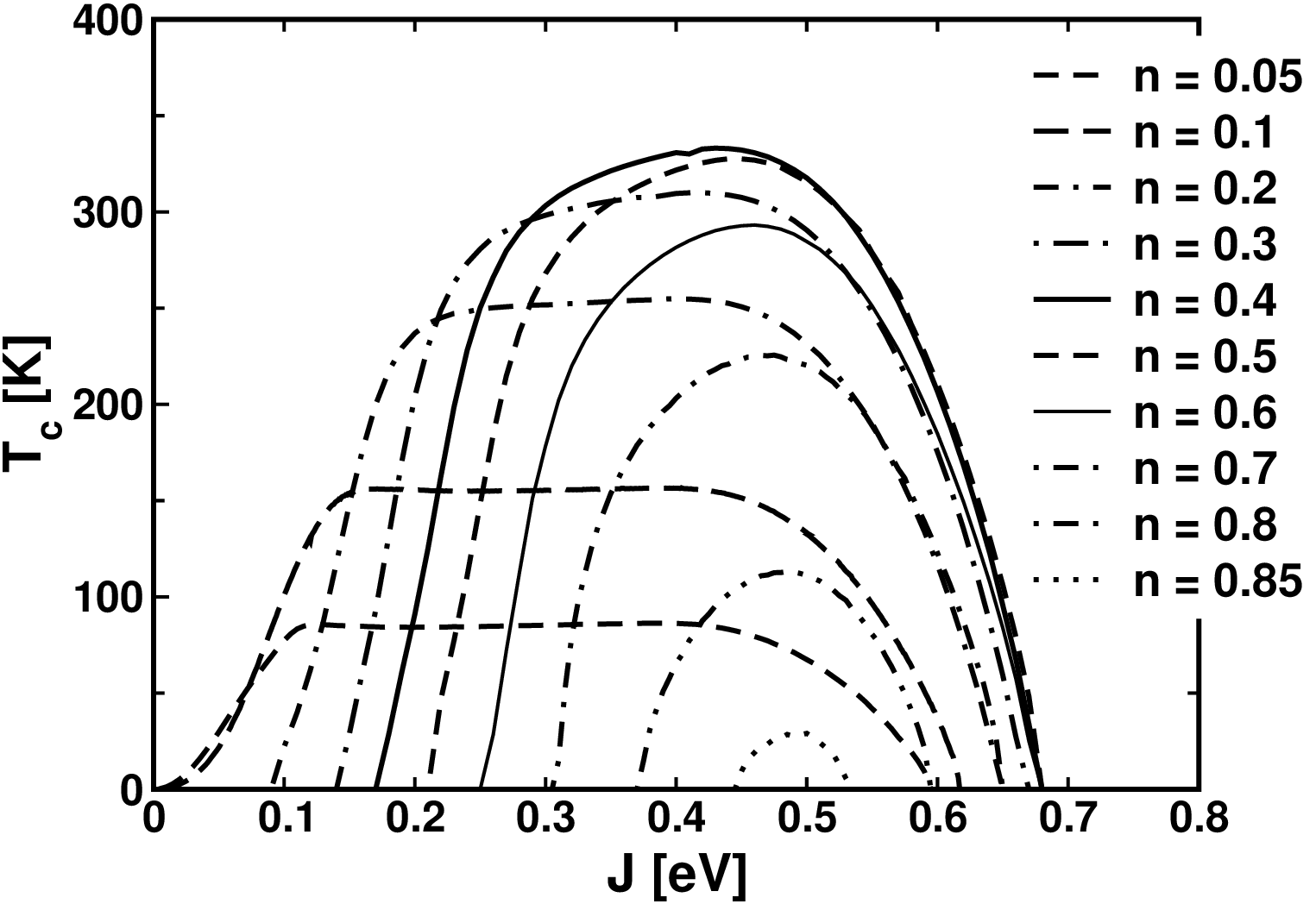}
\caption{Curie temperature as a function of the interband exchange
  coupling $J$ for various band occupations $n$ ($S=7/2$, s.c.~lattice,
  $W=1$ eV).}
\label{sn:fig:Tc_J}
\end{figure}

The Curie temperature shows a remarkable $J$ dependence, which is plotted
in Fig.~\ref{sn:fig:Tc_J} for various band occupations. According to
conventional RKKY theory one would expect a monotonic increase of
$T_{\textrm{C}}$ with increasing $J$. A mean-field evaluation of the effective Heisenberg
model would lead to $T_{\textrm{C}}\sim J^2$.
Our theory becomes identical to conventional RKKY theory for small
$J$. However, because of the random-phase-approximation (RPA) treatment 
of the Heisenberg operator
[Eq.~(\ref{sn:eq:Hf})], there are deviations of $T_{\textrm{C}}$ from the
$J^2$ behavior even for small $J$.
A more important and very characteristic feature of our modified RKKY
approach is the appearence of a critical coupling $J_c$ for band
occupations $n\ge 0.13$, below which no ferromagnetism occurs. This
occurs for band fillings, for which conventional RKKY theory fails to
yield a ferromagnetic RPA solution
(Fig.~\ref{sn:fig:Tc_n_Conv}). Therefore the existence of $J_c$ does not 
disprove the agreement of conventional and modified RKKY theory for small $J$.
After a steep increase as function of $J$, $T_{\textrm{C}}$ runs into a
kind of saturation, where the plateau becomes smaller the larger the
bandfilling $n$. With a further increase of $J$, the Curie temperature
again decreases and disappears at an upper critical $J$ value. The
upper critical value of $J$ coincides with the point where the stiffness
constant $D$ of the spin-wave dispersion [Eq.~(\ref{sn:eq:MagnonEnergies})]
($E(\vec{q})\approx D\:\vec{q}^2$ near the $\Gamma$ point)
becomes negative due to a respective behavior of the dominating
effective exchange integrals $J_1, J_2, J_3$ (see inset in
Fig.~\ref{sn:fig:Jij_J} 
and Eq.~(\ref{sn:eq:Tc})). It can be suspected that for
larger $J$ the local-moment system becomes
antiferromagnetic. Furthermore, it cannot be excluded that for still
larger $J$'s the system returns to ferromagnetism. These features are far
beyond conventional RKKY theory and due to higher order polarization
terms, which come into play via the full Green function
$G_{\vec{k}+\vec{q}\sigma}(E-\mu)$ in the expression (\ref{sn:eq:Jq})
for the effective exchange integral $\hat J(\vec{q})$. The manganites
are often treated as ferromagnets with $J\rightarrow\infty$ in the
KLM\cite{FUR96}. According to Fig.~\ref{sn:fig:Tc_J} this must be
reexamined. For $J\approx W$ ferromagnetism is unlikely in the KLM.
\begin{figure}[h]
\includegraphics[width=0.4\textwidth]{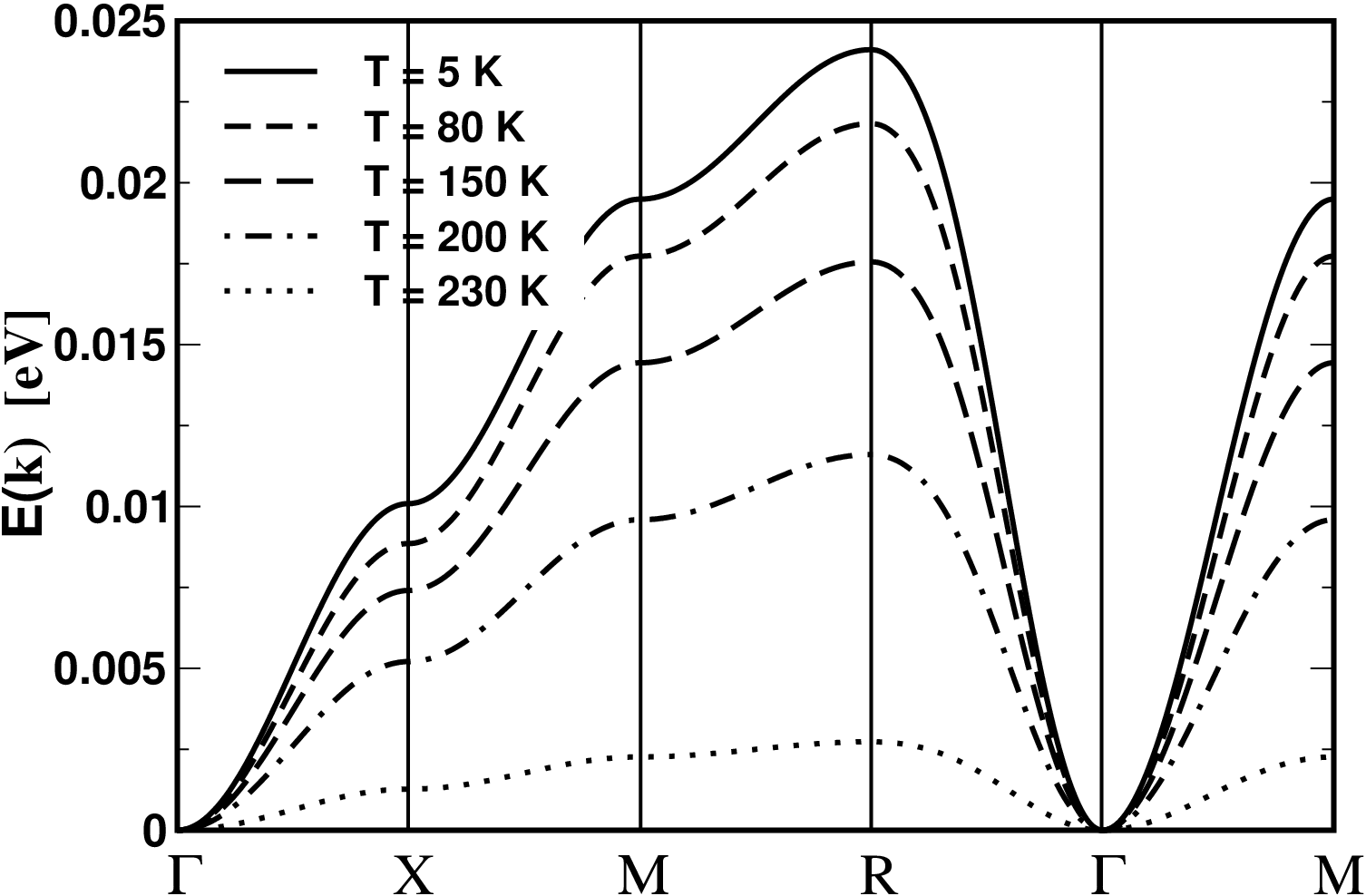}
\caption{Spin-wave dispersion of the ferromagnetic Kondo-lattice model
         as a function of the wave vector for different temperatures
         $T$. Parameters: $J=0.2$ eV, $n=0.2$, $S=7/2$, s.c.~lattice, 
         $W=1eV$. The self-consistently calculated Curie temperature is
         $T_{\textrm{C}}=232$ K.}
\label{sn:fig:MagnonT}
\end{figure}

The spin-wave dispersion, plotted in Fig.~\ref{sn:fig:MagnonT} for the main
symmetry direction, obtains, via the effective exchange integrals a distinct
temperature dependence. Upon heating the dispersion relation uniformly
softens, disappearing above $T_{\textrm{C}}$.
This agrees qualitatively
with the a neutron-scattering study of the spin dynamics in the manganite
Pr$_{0.63}$Sr$_{0.37}$MnO$_{3}$\cite{HDC98}. However, we do not share the
view of the authors of Ref.~\onlinecite{HDC98}, i.e., that the unexcepted
results rule out a simple Heisenberg Hamiltonian. In our opinion the
Heisenberg model works as long as the exchange integrals are
renormalized in a proper way by the conduction-electron self-energy.
\begin{figure}[h]
\includegraphics[width=0.4\textwidth]{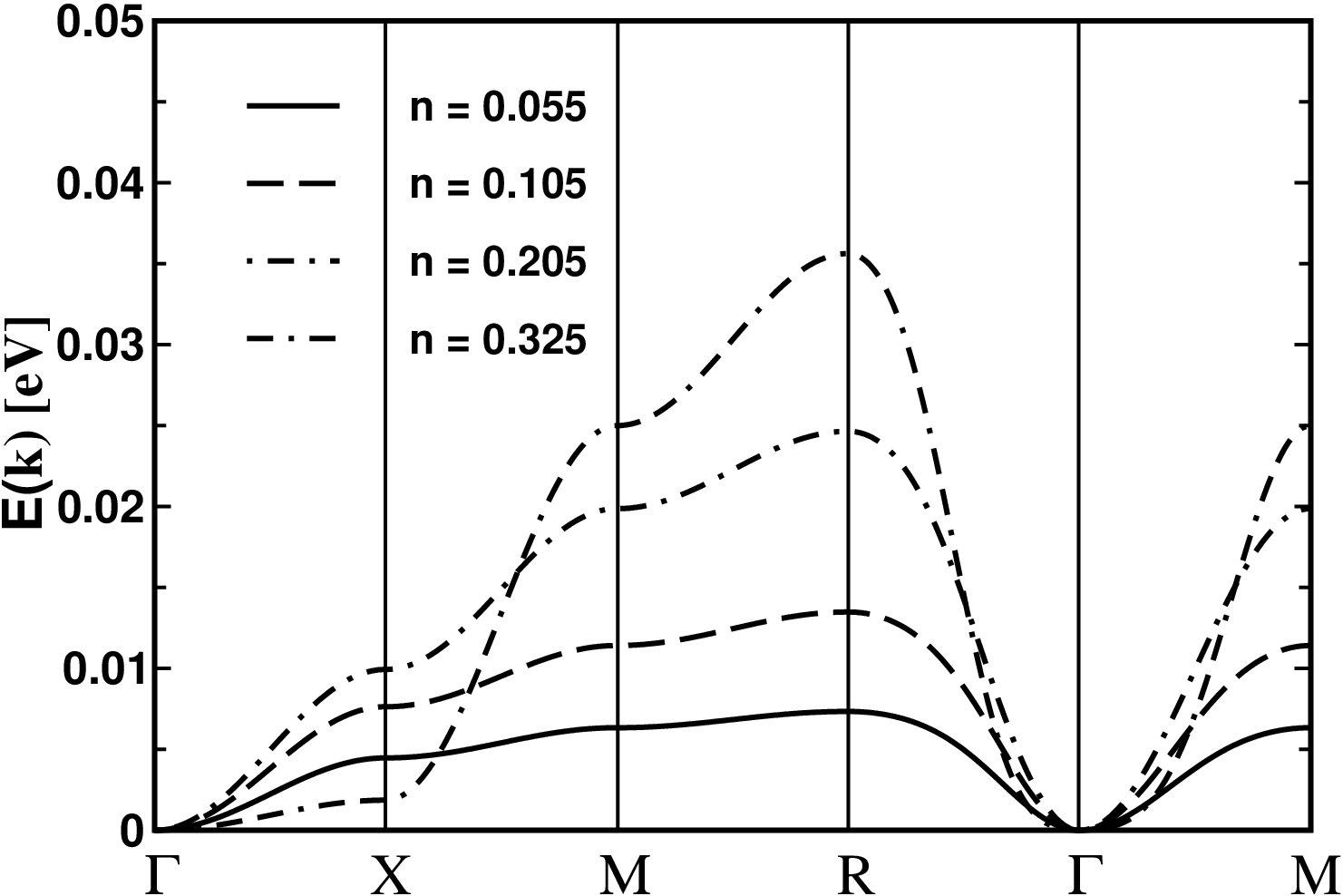}
\caption{The same as in Fig.~\ref{sn:fig:MagnonT}, but for different band
         occupations at $T=0$ K.}
\label{sn:fig:Magnon_na}
\end{figure}

\begin{figure}[h]
\includegraphics[width=0.4\textwidth]{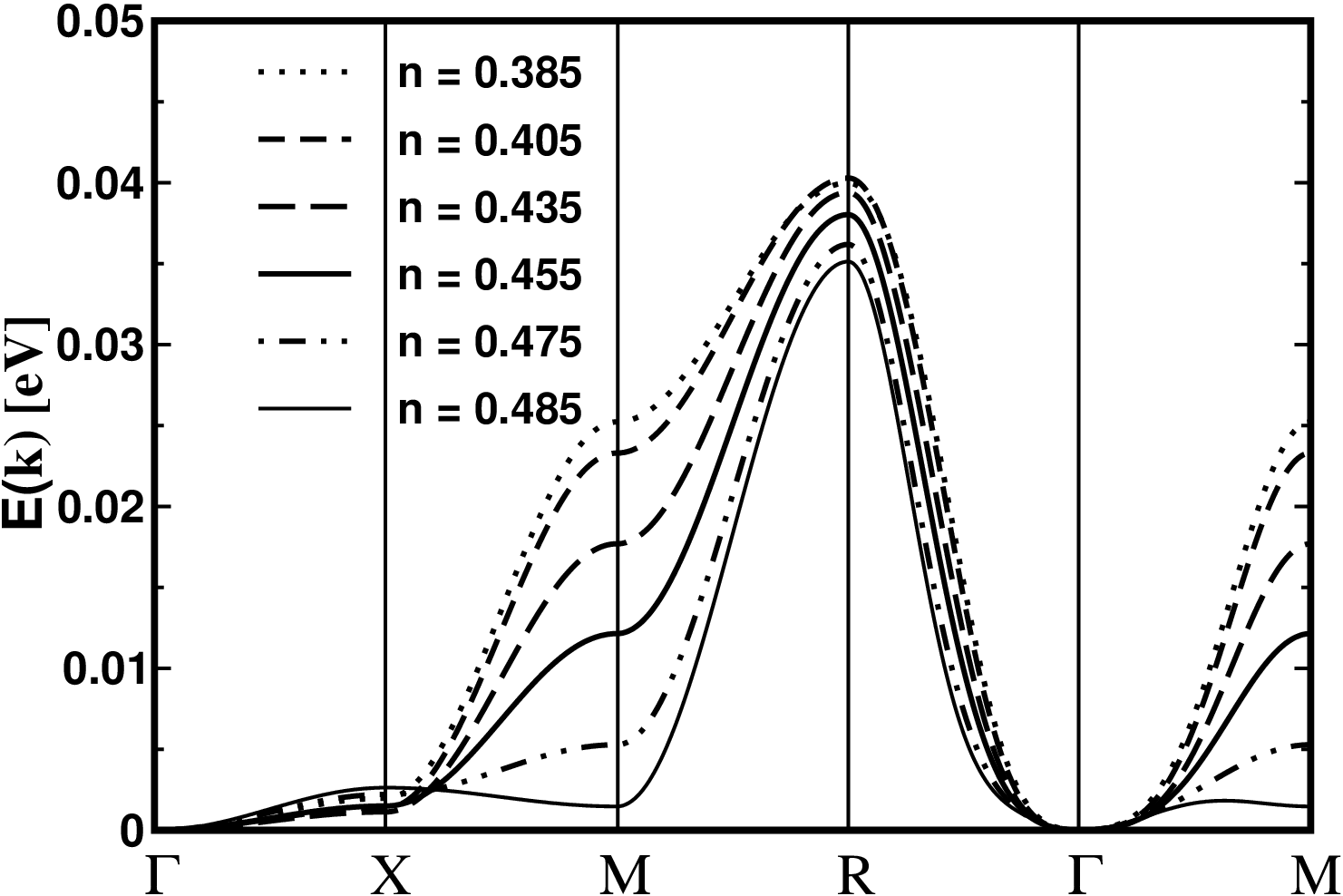}
\caption{The same as in Fig.~\ref{sn:fig:Magnon_na}, but for higher band 
         occupations.}
\label{sn:fig:Magnon_nb}
\end{figure}

Changing the doping in the III-V based diluted magnetic semiconductors
such as Ga$_{1-x}$Mn$_x$As, or in manganites like
La$_{1-x}$Ca$_x$MnO$_3$, does alter the band occupation and therewith the
physics of these materials. It is therefore worthwhile to inspect the
$n$ dependence of the spin-wave dispersion. This is done in
Fig.~\ref{sn:fig:Magnon_na} for low bandoccupations and in Fig.~\ref{sn:fig:Magnon_nb}
for higher band occupations for a model system with $J=0.2$ eV. 
For low densities
(Fig.~\ref{sn:fig:Magnon_na}) we observe a strengthening of the magnon
dispersion with increasing band filling, while the opposite is true for
stronger fillings (Fig.~\ref{sn:fig:Magnon_nb}). This is in accordance
with the $T_{\textrm{C}}$ behavior of Fig.~\ref{sn:fig:Tc_n}. The band
occupation $n=0.485$ is very close to the upper critical point of the
$J=0.2$ eV curve in Fig.~\ref{sn:fig:Tc_n}. The corresponding spin-wave
dispersion in Fig.~\ref{sn:fig:Magnon_nb} shows a strong softening.
A slightly higher $n$ leads to parts of the Brillouin zone with negative
magnon energies, i.e.~the ferromagnetic ground state becomes
unstable. The stiffness $D$, defined by $E(\vec{q})\approx D\:\vec{q}^2$
for $\vec{q}\rightarrow \vec{0}$, becomes negative. The points where
$D=D(n,J,T=0)= 0^+$ exactly coincide in our theory with those for
which $T_{\textrm{C}}=T_{\textrm{C}}(n,J)\rightarrow 0$, directly
derived from the local-moment magnetization $\langle
S^z\rangle$. Qualitatively the same $n$ dependence of the spin-wave
dispersion is reported in Ref.~\onlinecite{WMH98}.

\begin{figure}[h]
\includegraphics[width=0.4\textwidth]{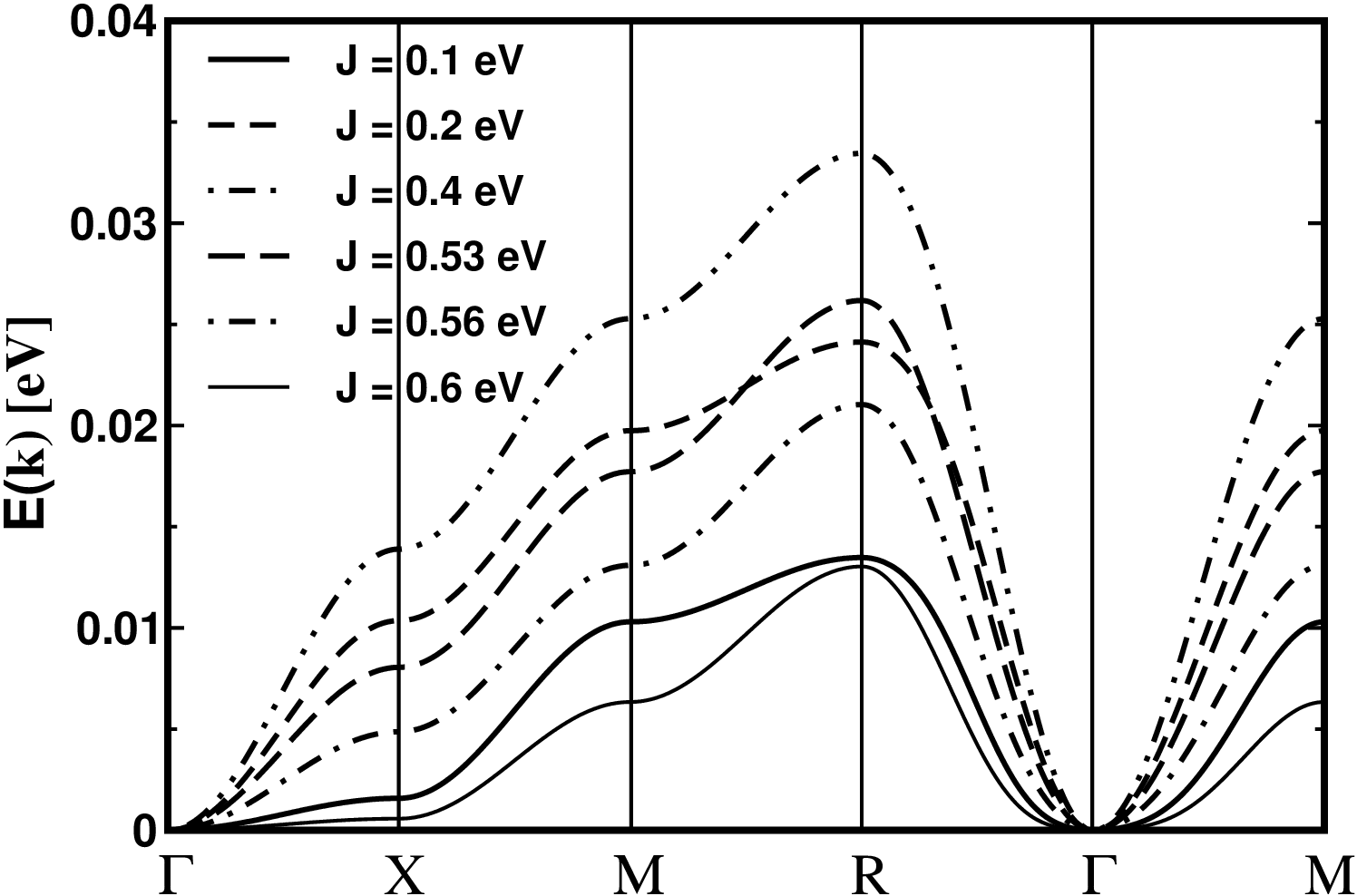}
\caption{The same as in Fig.~\ref{sn:fig:MagnonT}, but for different interband
         exchange couplings $J$ at $T=0$ K.}
\label{sn:fig:Magnon_J}
\end{figure}

An example for the $J$ dependence of the spin-wave dispersion is
presented in Fig.~\ref{sn:fig:Magnon_J} ($n=0.2$, $T=0$ K,
$S=\frac{7}{2}$). Again we observe a strengthening of the dispersion
with $J$ when $T_{\textrm{C}}$ increases simultaneously, and a
softening, when $T_{\textrm{C}}$ decreases (see Fig.~\ref{sn:fig:Tc_J}).
\section{Summary}
We have evaluated an approximate but self-consistent theory for a
local-moment ferromagnet described within the framework of the
ferromagnetic Kondo-lattice model. Candidates for this model are
magnetic semiconductors (EuO, EuS), diluted magnetic semiconductors
(Ga$_{1-x}$Mn$_x$As), magnetic metals (Gd, Dy, Tb) and CMR materials
(La$_{1-x}$Ca$_{x}$MnO$_3$). We have used a previously developed Green
function technique\cite{NRM97} for a detailed investigation of the
magnetic properties of the exchange-coupled local moment-itinerant
electron system. An extended RKKY mechanism has been worked out to derive
magnetic phase diagrams as well as spin-wave dispersions. The latter
show strong dependencies on temperature (uniform softening for
$T\rightarrow T_{\textrm{C}}$), on the band occupation $n$, and on the
exchange coupling $J$. These dependencies are due to the respective
behaviors of the effective exchange integrals between the localized
moments. The effective exchange integrals are found by mapping the
interband exchange ($s-f$) coupling, characteristic of the KLM, to an
effective Heisenberg model. They turn out to be functions of the
electronic self-energy, being therewith strongly temperature and
electron density dependent. For small $J$ our theory reproduces the
results of the well-known conventional RKKY treatment, which represents
second-order perturbation theory to the KLM. Already for very moderate
$J$, however, substantial deviations appear, when higher-order terms of
the induced conduction-electron spin polarization bring their influence
to bear.

For the future the presented approach has to be extended to
antiferromagnetic moment configurations in order to complete the phase
diagram. Furthermore, Coulomb correlations have to be introduced for
the conduction-electron subsystem. Their neglect in the original KLM
seems to be only poorly justified.

It is further intended to apply the modified RKKY procedure to the
antiferromagnetic KLM ($J<0$) in order to investigate the interplay
between ''Kondo screening'' and RKKY. This, however, requires a serious
check whether or not the MCDA in its present form\cite{NRM97} is able to
account for the subtle low-temperature (Kondo) physics.

\end{document}